\documentclass{article}
\usepackage{graphicx}
\usepackage{mathtools}
\usepackage{amsmath}
\usepackage{amssymb}
\usepackage{dsfont}
\usepackage{bbold}
\usepackage{siunitx}
\usepackage[left=2cm, right=2cm, top=2cm, bottom=2cm]{geometry}
\usepackage{caption}
\usepackage{subcaption}
\usepackage{float}

\usepackage{xargs}

\usepackage{xcolor}

\usepackage[backend=biber, sorting=none, style=numeric-comp, url=false, doi=false, isbn=false, eprint=false]{biblatex}
\title{The scaling behaviour of localised and extended states in one-dimensional tight-binding models with disorder}
\author{Luca Schaefer \and Barbara Drossel}
\date{Institute for condensed matter physics, Technische Universit\"at Darmstadt, Hochschulstraße 6, 64289 Darmstadt}

\begin{document}

\maketitle

\begin{abstract}
We investigate two one-dimensional tight-binding models with disorder that have extended states at zero energy. We use exact and partial diagonalisation of the Hamiltonian to obtain the eigenmodes and the associated participation ratios, and the transfer-matrix method to
determine the localisation length. The first model has
no on-site disorder, but random couplings. While the participation ratio remains finite at zero energy, the localisation length diverges logarithmically as the energy goes to zero. We provide an intuitive derivation of this logarithmic divergence based on the weak coupling of the two sublattices. The second model has a conserved quantity as the row sums of the Hamiltonian are zero. This model can be represented as a harmonic chain with random couplings, or as a diffusion model on a lattice with random links. 
We find, in agreement with existing analytical
calculations, that the number of system-spanning eigenmodes increases proportionally to the square root of the system size, and we related this
power law to other power laws that characterise the scaling behaviour of the eigenmodes, the participation ratio, the localisation length, and their dependence on energy and system size. When disorder is so strong that the smallest hopping terms can be arbitrarily close to zero, all these power laws change, and we show a crossover between the two scaling regimes. All these results are explained by intuitive arguments based on scaling. 
\end{abstract}

\section{Introduction}

The diffusion of particles in porous media, the dispersion of biological species on a network of habitats, and the motion of a tightly-bound electron in a randomly doped semiconductor, can all be described with a similar mathematical formalism. In the language of quantum mechanics, the corresponding model is the discretised Schrödinger equation that results from the tight-binding model, with added site and bond disorder.

In his seminal paper of 1958, Anderson showed that site disorder in the one-dimensional tight-binding model leads to localisation of electrons and thus renders the system an insulator~\cite{anderson_absence_1958}. Since  then, publications on this topic have never ceased. Anderson's findings were backed up by computer simulations, analytical calculations and experiments~\cite{mott_theory_1961, kramer_localization_1993}.
The model was generalised to also include bond disorder, i.e., off-diagonal disorder in the tight-binding Hamiltonian, and it was studied in various dimensions, using computer simulations, analytical calculations and experiments~\cite{kramer_localization_1993, ludlamDisorderInducedVibrationalLocalization2003, ludlamNumericalAnalysisVibrational2001a}.
Scaling arguments based on one-parameter scaling show that in one and two dimensions all states are localised for any disorder strength, whereas in three dimensions a metal-insulator transition is observed~\cite{abrahams_scaling_1979}.

However, there are exceptions to the rule that disorder leads to localisation of all einenmodes in one-dimensional systems. Special cases of the model where not all states are localised even in one dimension, were identified. If the disorder resides only in the hopping elements and not on the onsite potentials, one obtains the random-coupling model (RCM). In this model, the state in the band center has infinite localisation length ~\cite{theodorouExtendedStatesOnedemensional1976,antoniouAbsenceAndersonTransition1977}.
However, Soukoulis and Economou argue that this state is nevertheless localised since its transmission goes to zero for infinite system size~\cite{soukoulisOffdiagonalDisorderOnedimensional1981}.
Inui and Trugman came to the conclusion that it is the bipartite structure of the lattice that leads to the infinite localisation length  in the band centre since the wave function on one sublattice behaves like a random walk~\cite{inuiUnusualPropertiesMidband1994}.
The importance of a bipartite lattice structure for obtaining an infinite localisation length was confirmed in two dimensions by Eilmes et al.~by switching from a square lattice to a honeycomb and a triangular lattice~\cite{eilmesExponentsLocalizationLengths2001a}.  By performing a numerical evaluation of the participation ratio and the localisation length, they also showed that adding even a small amount of diagonal disorder makes the localisation length in the band centre finite~\cite{eilmes_two-dimensional_1998}.

A particularly interesting case is obtained by taking the on-site potential as the negative sum of the adjacent hopping elements, where the Hamiltonian becomes a discretised Laplacian, implying that the model has a conserved quantity. We call  this model the diffusion model (DM). It is relevant for  diffusion processes on lattices with random couplings, and it also describes a (classical) chain of equal masses coupled by harmonic springs of random coupling strengths.
Dean~\cite{dean_vibrations_1964, dean_vibrational_1972} investigated this model numerically and analytically, finding that the localisation length of eigenvectors increases with decreasing energy. 
Using a transfer-matrix calculation in the three-dimensional model, Pinski et al.~found extended states at $E=0$~\cite{pinskiAndersonUniversalityModel2012}. This result holds also when  masses and spring constants may be negative, a case that applies to acoustic metamaterials~\cite{pinskiAndersonUniversalityModel2012}.
Ludlam et al.~conducted a multifractal analysis of the vibrational modes in two- and three-dimensional lattices  with disorder in the spring constants~\cite{ludlamNumericalAnalysisVibrational2001a,ludlamDisorderInducedVibrationalLocalization2003}. They found that states in the band tails are localised and the other states are pre-localised, i.e. multifractal with an extension larger than the studied system sizes.
The scaling behaviour of this model was also investigated by Dunlap et al.~\cite{dunlap_absence_1989}, who found superdiffusive propagation and calculated that the number of  system-spanning modes in systems of linear dimension $L$ increases as $\sqrt{L}$.  
The strength of disorder is usually reported to have only a quantitative influence, but not a qualitative one. However, couplings close to zero represent barriers to diffusion in one-dimensional diffusion processes. For this reason, some authors introduce a lower cutoff larger than zero \cite{dunlap_absence_1989,dean_vibrational_1972}.

We are not aware of a systematic investigation of the behaviour of the model as the cutoff approaches zero. While mathematically elaborate calculations were presented for the one-dimensional model, we miss an intuitive and systematic understanding of the scaling laws that occur in the DM and of the relation between them.  

In this paper, we will revisit the one-dimensional version of the RCM and the DM and will provide an in-depth intuitive understanding of these models and their various scaling laws. To this purpose, we investigate the eigenmodes of these models that are obtained by diagonalising the Hamiltonian, and we evaluate in particular the participation ratio, which is a measure of the extension of the eigenmodes. We will also evaluate the localisation length that results from the transfer matrix method. For the DM, we obtain a variety of power laws and scaling collapses that show how the localised and system-spanning modes depend on energy, system size, and the lower cutoff of the  coupling strenghts. We show that the case where the lower cutoff is zero is described by different power laws than the case of a nonzero cutoff, and we show the crossover between these two cases as the system size increases. For all our results, we provide intuitive explanations and scaling arguments that provide a deeper understanding of the one-dimensional DM. 

While in the DM the participation ratio and the localisation length scale in the same way with energy and disorder, we find that the RCM behaves differently: The localisation length of the RCM diverges as $\ln E$ as the energy approaches zero, while the participation ratio approaches a constant value. We will explain the logarithmic divergence of the localisation length based on the weak coupling between the two sublattices for small energies. We ascribe the reason why the participation ratio does not show this divergence to the exponentially sharp maxima of the eigenmodes.

\section{Models}
The starting point of our investigation is the discrete tight-binding Schrödinger equation in one dimension (Eq.~\eqref{eq:schroedinger_equation}).
\begin{align}\label{eq:schroedinger_equation}
    E\psi_n = \epsilon_n\psi_n - t_{n,n+1}\psi_{n+1} - t_{n,n-1}\psi_{n-1}
\end{align}
This equation is obtained from the tight-binding model, which describes non-interacting electrons in a lattice, with the attraction between lattice ions and electrons being so strong that electrons are localised in the vicinity of ions. This equation can also be obtained directly by discretising the one-dimensional Schr\"odinger equation. The $t_{ij}$ are the hopping matrix elements which result from the kinetic energy term and characterise the transition amplitudes of an electron from one potential well to another. These and the on-site energies $\epsilon_n$ may be random variables in a system with disorder.

We distinguish three different models depending on the choice of $\epsilon_n$ and $t_{ij}$.
The Anderson model (AM)~\cite{anderson_absence_1958} is obtained for $t_{ij}\equiv 1$ and random $\epsilon_n$, which are often taken from a bounded symmetric distribution on an interval $\left[-\frac{W}{2},\frac{W}{2}\right],~W\in\mathbb{R}$.
The second model, which we call random-coupling model (RCM), has $\epsilon_n=0$ and $t_{ij}\in[c-\frac{w}{2},c+\frac{w}{2}],~c,w\in\mathbb{R}$~\cite{soukoulisOffdiagonalDisorderOnedimensional1981,eilmes_two-dimensional_1998}. 
The third model is the diffusion model (DM) with $t_{ij}$ as in the RCM but $\epsilon_n = t_{n,n-1}+t_{n,n+1}$~\cite{pinskiAndersonUniversalityModel2012}.  Without loss of generality, we set $c$ to 1 in both models.
For $w\leq2c$, the negative of the right-hand side of Eq.~\eqref{eq:schroedinger_equation} is the one-dimensional version of the positive-semidefinite graph Laplacian, which is the generalisation of the discretised Laplace operator to random couplings. The time-dependent version of this model (with $i\hbar \partial\psi_n/\partial t$ on the left-hand side instead of $E\psi_n$) has a conserved quantity $\sum_n \psi_n$. While it may be difficult to find a motivation for this model within the framework of the quantum-mechanical description of electrons in disordered lattices (except from artificial construction of such a system), there are various applications of this model in other fields of physics. In classical mechanics, the model describes a system of masses coupled by springs,
\begin{equation}
    m_n\ddot x_n = -D_{n,n+1}(x_{n}-x_{n+1}) - D_{n,n-1}(x_n-x_{n-1})\, ,
\end{equation}
where the $x_n$ represent the deviations from the equilibrium positions and $D_{n,n+1}$ is the spring constant between the masses $m_n$ and $m_{n+1}$. This system of equations can be solved with the ansatz $x_n(t)=x_n(0)e^{-i\omega t}$, leading to 
\begin{equation}
    m_n \omega^2 x_n = D_{n,n+1}(x_{n}-x_{n+1}) + D_{n,n-1}(x_n-x_{n-1})\, .
\end{equation}
This is an equation for the oscillating eigenmodes of the system, with $\omega$ being the oscillation frequency. 
When masses differ from each other, this is the general model Eq.~\eqref{eq:schroedinger_equation}. When they are identical, it is the diffusion model.

Another, very broad, field of applications of the diffusion model is that of diffusion processes between discrete sites (hence our name for the model). In this case, the $\psi_n$ are replaced by the amounts $c_n$ of the substances that diffuse between sites, which may be, for instance, molecules, heat, or individuals of a population, depending on the application. In ecological applications, the sites represent habitats. The time evolution of such systems is described by the equations
\begin{equation}
    \frac{\partial c_n}{\partial t} = -\sum_{j\in\mathrm{NN}(n)}D_{nj}(c_n-c_j) \label{DMt}
\end{equation}
with the sum being taken over the nearest neighbours of $n$, which are $n-1$ and $n+1$ for a one-dimensional chain of habitats. The $D_{nj}$ are now the rates of transfer between sites or habitats $n$ and $j$. The total amount $\sum_n c_n$ is a conserved quantity. With the ansatz $c_n(t)=x_n(0)e^{-t/\tau}+ c_n^{\mathrm{eq}}$, one obtains an equation for the relaxing eigenmodes,
\begin{equation}
    \frac{c_n}{\tau} = \sum_{j\in\mathrm{NN}(n)}D_{ij}(c_n-c_j)\, , \label{DM2}
\end{equation}
with the relaxation constant $1/\tau$. This equation is again identical in form to the quantum-mechanical version of the diffusion model. 
The case $w=2c$ is special for this model, as the couplings $t_{ij}$  can become very small  and diffusion through such a link becomes arbitrarily slow;  we will see below that this leads to qualitatively different results from the case $w < 2c$. 

The general model Eq.~\eqref{eq:schroedinger_equation} is obtained 
by adding source or sink terms  $ + u_n c_n$ on the right-hand side of Eq.~\eqref{DMt}.

\section{Methods}
In order to quantify the localisation behaviour, we use the participation ratio of the eigenstates of the Hamiltonian and the localisation length obtained from the transfer matrix method (TMM).
For all computations, analyses, and figures, we used the Julia programming language (version 1.9.1).
\subsection{Evaluating eigenfunctions obtained from the diagonalisation of the Hamiltonian}
The participation ratio ($P$) of an eigenstate $\vec{\psi}$ is defined as
\begin{align}\label{eq:participation_ratio}
    P\coloneqq\frac{\left(\sum_i^L\psi_i^2\right)^2}{\sum_i^L\psi_i^4}\, .
\end{align}
It is equal to the system length $L$ for a state that is equally distributed over all sites (extended) and $1$ for a state that has only one nonzero component (localised). 
For a sine function (which will become relevant further below), the participation ratio is $P=2/3 L$.
We will evaluate the probability density of $P$  and the dependence of $P$ on the energy eigenvalues $E$ for different system sizes.
For this, a diagonalisation of the Hamiltonian is performed with the Julia library  \texttt{LinearAlgebra}.
The Hamiltonian matrix in the basis of position eigenstates is
\begin{align}
    \begin{pmatrix}
        \epsilon_1 & -t_{1,2} &  &  & -t_{1,L}\\
        -t_{1,2} & \epsilon_2 & -t_{2,3} &  &  \\
         & -t_{2,3} & \epsilon_3 & t_{3,4} &  \\
         & & &  & \\
         & & & \ddots & \\
         & & & & \\
        -t_{1,L} &  &  & -t_{L-1,L} & \epsilon_L
    \end{pmatrix} ~,
\end{align}
i.e. a symmetric tridiagonal matrix with two additional nonzero elements at $1L$ and $L1$, which are due to periodic boundary conditions.
With this method, we are able to diagonalise matrices up to $20480\times20480$.
For larger systems, we resort to sparse representations of $H$ and the partial diagonalisation of symmetric matrices (\texttt{eigs} in \texttt{LinearAlgebra}).
In our case, this means that we only computed the 10\% of the eigenvalues (and their eigenvectors) closest to zero for systems up to $81920\times 81920$.

\subsection{Evaluating the localisation length using the Transfer-Matrix Method (TMM)}
Equation~\eqref{eq:schroedinger_equation} can be rewritten as
\begin{align}
    \begin{pmatrix}
        \psi_{n+1}\\
        \psi_n
    \end{pmatrix} &= 
    \begin{pmatrix}
        \frac{\epsilon_n-E}{t_{n,n+1}} & -\frac{t_{n,n-1}}{t_{n,n+1}}\\
        1 & 0
    \end{pmatrix}
    \cdot\begin{pmatrix}
        \psi_n\\
        \psi_{n-1}
    \end{pmatrix} \label{eqTM}\\
    \begin{pmatrix}
        \psi_{L+1}\\
        \psi_L
    \end{pmatrix}&=\underbrace{\prod_{i=1}^L
    \begin{pmatrix}
        \frac{\epsilon_i-E}{t_{i,i+1}} & -\frac{t_{i,i-1}}{t_{i,i+1}}\\
        1 & 0
    \end{pmatrix}}_{\eqcolon T_L}
    \cdot\begin{pmatrix}
        \psi_1\\
        \psi_{0}
    \end{pmatrix}~.
\end{align}
The wave function can thus be constructed iteratively by performing a multiplication of transfer matrices.

In the absence of disorder, we obtain an analytical expression for the product of transfer matrices, 
\begin{align}\label{eq:TMM_analytic}
    \begin{pmatrix}
        \epsilon & -1\\
        1 & 0
    \end{pmatrix}^L
    &=\frac{2}{\sqrt{4-\epsilon^2}}
    \begin{pmatrix}
        \sin\left[(L+1)\arctan\left(\sqrt{\frac{4}{\epsilon^2}-1}\right)\right] & 
        -\sin\left[L\arctan\left(\sqrt{\frac{4}{\epsilon^2}-1}\right)\right]\\
        \sin\left[L\arctan\left(\sqrt{\frac{4}{\epsilon^2}-1}\right)\right] & 
        -\sin\left[(L-1)\arctan\left(\sqrt{\frac{4}{\epsilon^2}-1}\right)\right]
    \end{pmatrix}~,
\end{align}
where $\epsilon= 2-E$ for the DM and $\epsilon = -E$ for the RCM.
For $\epsilon\to 0$, the arctan function on the right-hand side goes to $\pi/2$. The transfer matrix is then a rotation by $\pi/2$ and the product matrix evaluates to a rotation by $(L\mod4)\cdot\pi/2$. This describes the eigenmodes with the shortest wavelength, which occur at the band edge ($E=2$) for the DM and at the band centre ($E=0$) for the RCM. 

For the RCM, the case $\epsilon=-E=0$ can be evaluated analytically also in the case with disorder, as the diagonal of the transfer matrix stays zero:
\begin{align}
    \prod_{n=1}^L  \begin{pmatrix}
        0 & -\frac{t_{n,n-1}}{t_{n,n+1}}\\
        1 & 0
    \end{pmatrix} &= 
    \begin{pmatrix}
        -\frac{t_{2,1}}{t_{2,3}} & 0\\
        0 & -\frac{t_{1,0}}{t_{1,2}}
    \end{pmatrix}
    \prod_{n=3}^L  \begin{pmatrix}
        0 & -\frac{t_{n,n-1}}{t_{n,n+1}}\\
        1 & 0
    \end{pmatrix}\\
    &= \begin{pmatrix}
        \prod_{n=1}^{L/2}\left(-\frac{t_{2n,2n-1}}{t_{2n,2n+1}}\right) & 0\\
        0 & \prod_{n=1}^{L/2}\left(-\frac{t_{2n-1,2n-2}}{t_{2n-1,2n-2}}\right)
    \end{pmatrix}\\
    &=
    \prod_{n=1}^{L/2}
    \begin{pmatrix}
        -\frac{t_{2n,2n-1}}{t_{2n,2n+1}} & 0\\
        0 & -\frac{t_{2n-1,2n-2}}{t_{2n-1,2n-2}}
    \end{pmatrix}~.
\end{align}
For the initial vector $\Vec{\psi}_0=(\cos(\varphi),\sin(\varphi))^T$, this gives the sequence
\begin{equation}
    \begin{pmatrix}-\frac{t_{0,1}}{t_{1,2}}\sin(\varphi)\\\cos(\varphi)\end{pmatrix}
    \to \begin{pmatrix}-\frac{t_{1,2}}{t_{2,3}}\cos(\varphi)\\-\frac{t_{0,1}}{t_{1,2}}\sin(\varphi)\end{pmatrix}
    \to \begin{pmatrix}\frac{t_{0,1}t_{2,3}}{t_{1,2}t_{3,4}}\sin(\varphi)\\-\frac{t_{1,2}}{t_{2,3}}\cos(\varphi)\end{pmatrix} 
    \to \begin{pmatrix}\frac{t_{1,2}t_{3,4}}{t_{2,3}t_{4,5}}\cos(\varphi)\\\frac{t_{0,1}t_{2,3}}{t_{1,2}t_{3,4}}\sin(\varphi)\end{pmatrix}  
    \to \dots
    \label{eq:independent}
\end{equation}
which is a superposition of two sequences for $\varphi=0$ and $\varphi = \pi/2$. 
The one for $\varphi=0$ is 
\begin{align}
    \begin{pmatrix}0\\1\end{pmatrix}
    \to \begin{pmatrix}-\frac{t_{1,2}}{t_{2,3}}\\0\end{pmatrix}
    \to \begin{pmatrix}0\\-\frac{t_{1,2}}{t_{2,3}}\end{pmatrix} 
    \to \begin{pmatrix}\frac{t_{1,2}t_{3,4}}{t_{2,3}t_{4,5}}\\0\end{pmatrix}  
    \to \dots
\end{align}
The logarithm of the absolute value of the nonzero entries of the $\psi_n$ can be written as a symmetric random walk 
\begin{align}\label{eq:RCM_RW}
    \log(t_{1,2}) - \log(t_{2,3}) + \log(t_{3,4}) - \log(t_{4,5}) + \dots
\end{align}

The localisation length is the inverse of the smallest Lyapunov exponent,
\begin{align}\label{defxi}    (\xi)^{-1}\coloneqq\lim_{L\to\infty}\frac{1}{L}\ln(\lVert T_L\vec{\psi}_0\rVert)~.
\end{align}

\subsection{Data handling}
We could perform a full diagonalisation for matrix sizes up to $L=20480$, before we had to resort to partial diagonalisation, determining only a tenth of all eigenvectors, which belong to the smallest absolute values of $E$.  
We scaled the ensemble sizes with the inverse system size such that the total number of eigenmodes is the same for all system sizes (tab.~\ref{tab:parameters_diagonalisation}).

\begin{table}[H]
    \centering
    \begin{tabular}{|c|c|c|c|}
    \hline
       $L$  & $w$ & ensemble size & \# eigenvectors\\
       \hline
      2560 & 1,2   & 800           & all\\
      5120 & 1,2   & 400           & all\\
      10240& 1,2   & 200           & all\\
      20480& 1,2   & 500           & all\\
      40960& 1,2   & 25            & 4096\\
      81920& 1,2   & 12            & 8192\\
             \hline
    \end{tabular}
    \caption{System sizes and sample sizes used for the full and partial diagonalisation of the Hamiltonian. }
    \label{tab:parameters_diagonalisation}
\end{table}

Due to the exponential growth of the matrix norm in the TMM, we normalised the vector after each multiplication and added the logarithm of the norm to the previous one.
\begin{align}
    \ln\left(\lVert \vec{\psi_L}\rVert\right) &= \ln\left(\lVert T_{L}\vec{\psi}_{L-1}\rVert\right) \\
    &= \ln\left(\lVert T_{L}\vec{\psi}_{L-1}\rVert\frac{\lVert \vec{\psi}_{L-1}\rVert\cdot\lVert \vec{\psi}_{L-2}\rVert\cdots\lVert \vec{\psi}_{1}\rVert}{\lVert \vec{\psi}_{L-1}\rVert\cdot\lVert \vec{\psi}_{L-2}\rVert\cdots\lVert \vec{\psi}_{1}\rVert}\right)\\
    &= \ln\left(\frac{\lVert T_{L}\vec{\psi}_{L-1}\rVert}{\lVert \vec{\psi}_{L-1}\rVert}\right) + \ln\left(\frac{\lVert T_{L-1}\vec{\psi}_{L-2}\rVert}{\lVert \vec{\psi}_{L-2}\rVert}\right) + \dots + \ln\left(\frac{\lVert T_{1}\vec{\psi}_{0}\rVert}{\lVert \vec{\psi}_{0}\rVert}\right)
\end{align}
We set $\vec{\psi}_0\equiv (1,0)^T$.

Since Lyapunov exponent converges for almost all values of $E$ (apart from $E=0$ in the RCM), we used a single realisation of disorder for each $E \neq 0$ and $w$.

\section{Results}
\subsection{Density of states (DOS)}
\begin{figure}[H]
     \centering
     \begin{subfigure}[b]{0.49\textwidth}
         \centering
         \includegraphics[width=\textwidth]{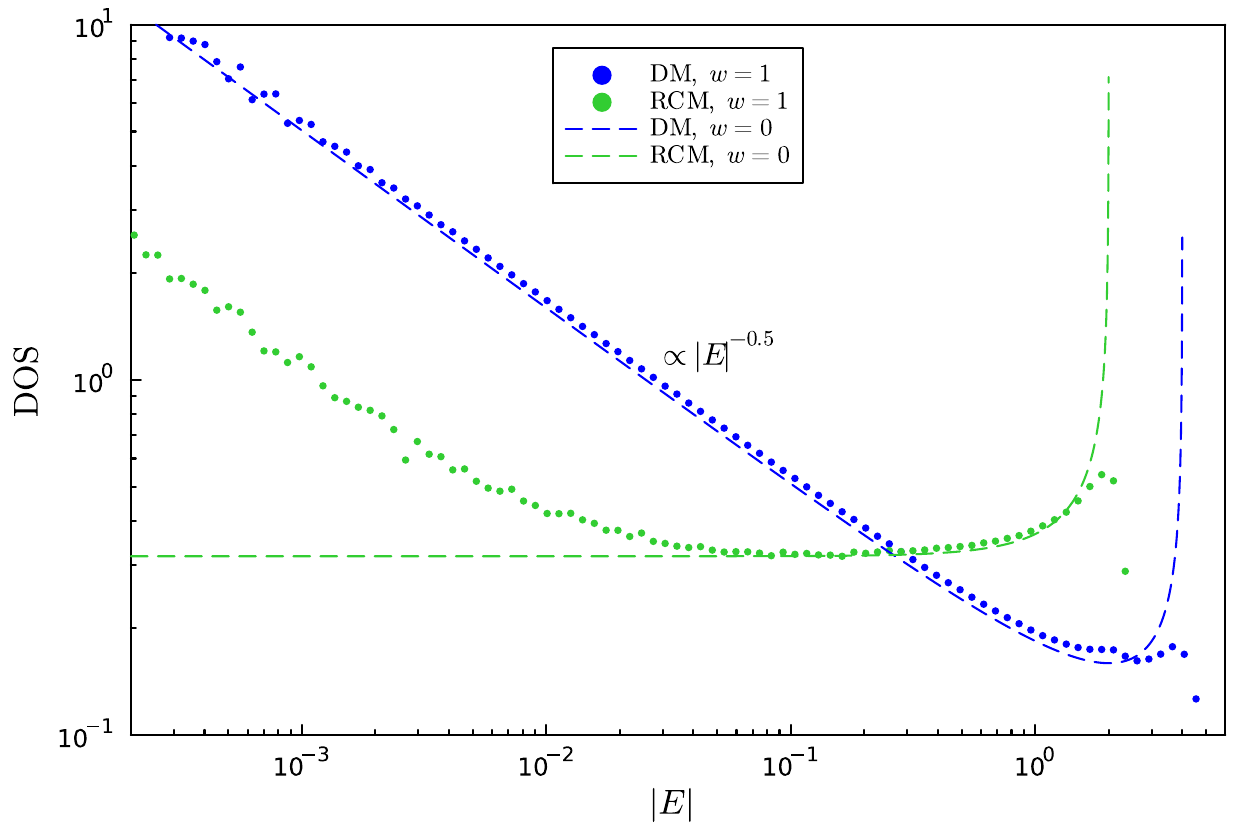}
         \caption{}
     \end{subfigure}
     \hfill
     \begin{subfigure}[b]{0.49\textwidth}
         \centering
         \includegraphics[width=\textwidth]{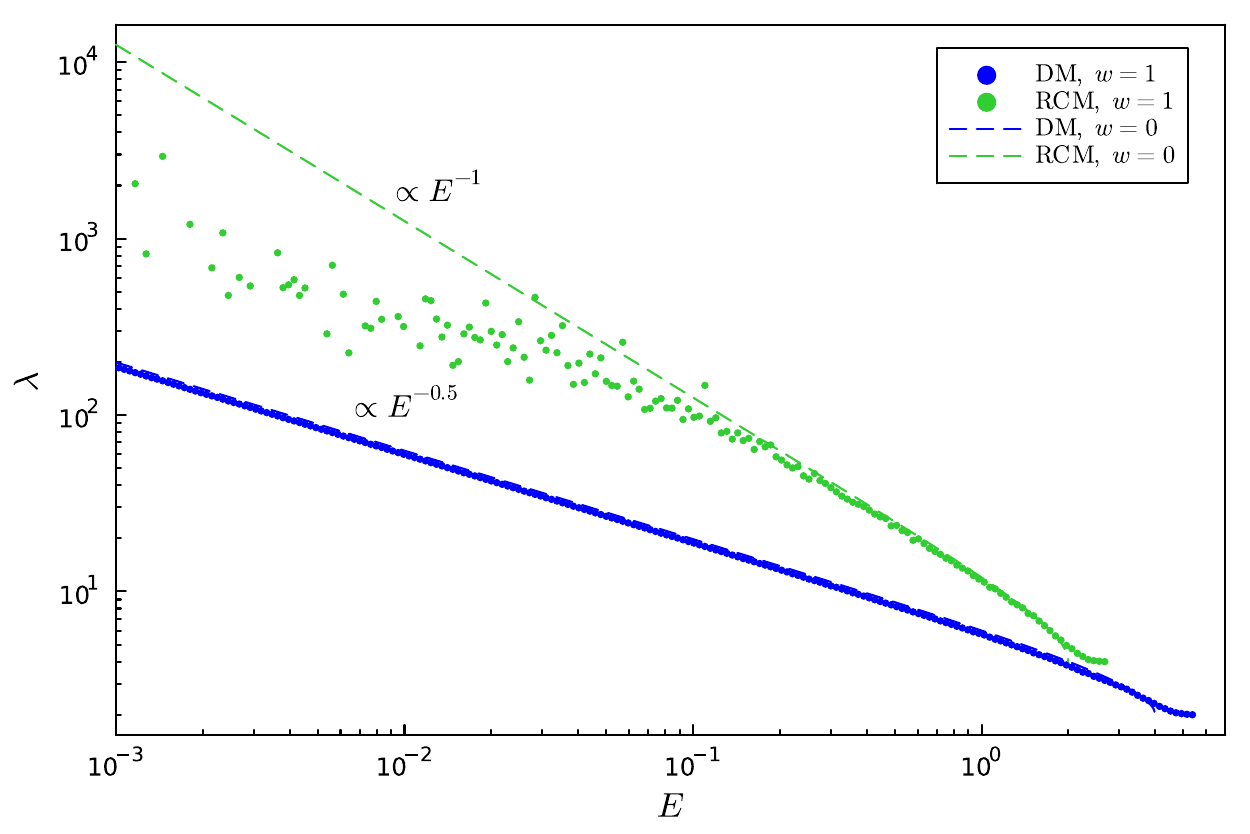}
         \caption{}
     \end{subfigure}
     \begin{subfigure}[b]{0.49\textwidth}
         \centering
         \includegraphics[width=\textwidth]{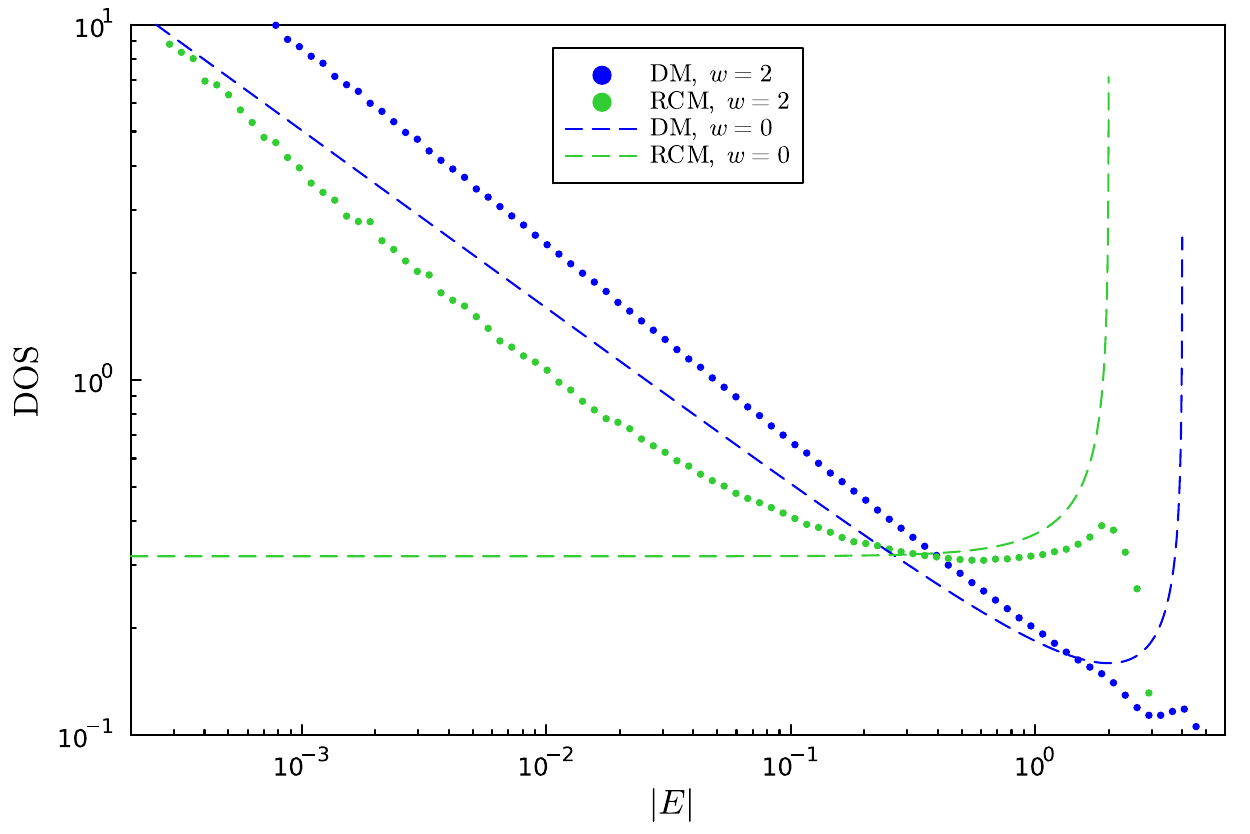}
         \caption{}
     \end{subfigure}
     \hfill
     \begin{subfigure}[b]{0.49\textwidth}
         \centering
         \includegraphics[width=\textwidth]{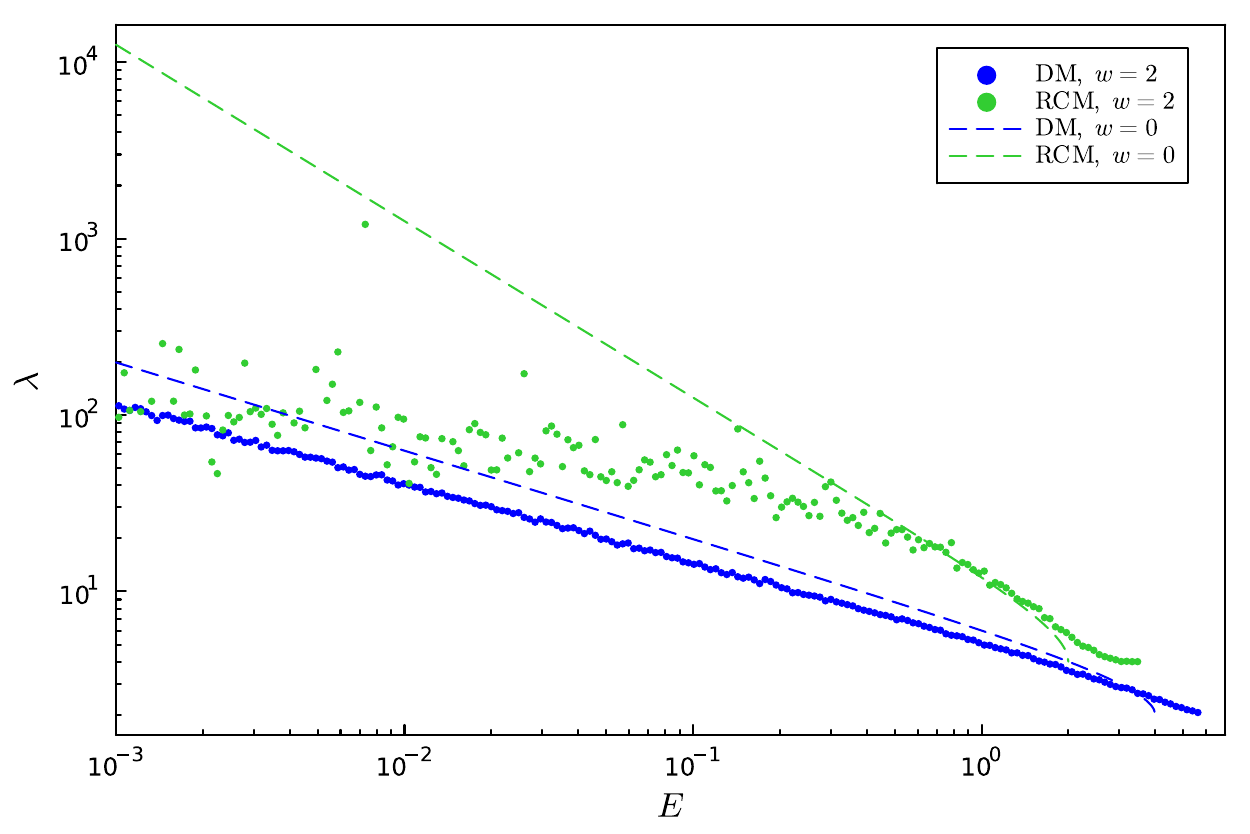}
         \caption{}
     \end{subfigure}
        \caption{Left: Density of states of the RCM and the DM for $L=20480$ and $w=1$ (a) and $w=2$ (c). The dashed lines correspond to disorder strength $w=0$.
        Right: Dependence between the energy and the wavelength of the dominant Fourier component for the RCM and DM ($L=20480$) for $w=1$ (b) and $w=2$ (d).   Wavelengths of $\lambda=L$ have been omitted. In the RCM, the wavelength $\lambda=4$ was also omitted, and only one sublattice was evaluated.  The dashed lines correspond to disorder strength $w=0$.} 
        \label{fig:DOS}
\end{figure}

The dispersion relation in the absence of disorder follows from Eq.~\eqref{eq:schroedinger_equation} by setting $t_{ij}=c$ and $\epsilon_n=0$ (for the RCM) or $2c$ (for the DM) and making the ansatz $\psi_n = \exp(ikn)$.
For the RCM we obtain
\begin{align}\label{eq:dispersion_RCM}
    E(k)&=-c\left(\exp(ik)-\exp(-ik)\right)\\
    &= -2c\cos(k)\, .\nonumber
\end{align}
For the DM, the energy scale is shifted by $2c$, and the dispersion relation becomes
\begin{align}\label{eq:dispersion_DM}
    E(k) &= 2c-2c\cos(k)\\
    &=2c\sin^2\left(\frac{k}{2}\right)\nonumber\\
    &=2c\sin^2\left(\frac{n \pi}{\lambda}\right)\nonumber\\
    &\overset{E\ll c}{\sim}\frac{1}{\lambda^2}~.
\end{align}
From the dispersion relation one obtains the density of states 
\begin{align}
\mathrm{DOS}(E) &= \left|\frac{\mathrm{d}k(E)}{\mathrm{d}E}\right| n(k)~,
\end{align}
where $n(k)=\frac{L}{2\pi}$ is the density of $k$ values in 1D.
Solving Eqs.~\eqref{eq:dispersion_RCM} and  \eqref{eq:dispersion_DM} for $k$ gives
\begin{align}
    k(E) = \arccos\left(-\frac{E}{2c}\right) \quad \hbox{ and } \quad  k(E) = \arccos\left(1-\frac{E}{2c}\right)\, .
\end{align}
The density of states of the two models in the absence of disorder is therefore 
\begin{align}\label{eq:DOS}
    \mathrm{DOS}_{\mathrm{RCM}}(E) &= \frac{L}{4\pi\sqrt{1-\frac{E^2}{4c^2}}} \overset{E\ll c}{\simeq} \frac L {4\pi}~,\\
    \mathrm{DOS}_{\mathrm{DM}}(E) &= \frac{L}{4\pi\sqrt{1-\frac{(2c-E)^2}{4c^2}}}\overset{E\ll c}{\propto} \frac{1}{\sqrt{E}}~.
\end{align}
Both DOS are shown in Fig.~\ref{fig:DOS} (dashed lines) together with the simulation data with a disorder strength $w=1$ (top row) and $w=2$ (bottom row). In this and all other figures, we have set $c=1$. 
The data is normalised, that is, DOS$_{\mathrm{RCM}}$ was multiplied by $4/L$, and DOS$_{\mathrm{DM}}$ by $2/L$.
Due to the shift of the energy axis by $2c=2$ between the two models, the curve of the RCM is symmetric with respect to $E=0$ with singularities at $\pm 2$, while the one for the DM is symmetric with respect to $E=2$, implying that one of the singularities  appears at zero energy in the DM.

In the presence of disorder ($w=1$ or 2), the singularity of the DOS at the upper band edge vanishes for both models~\cite{schirmacherHarmonicVibrationalExcitations1998}; see Fig.~\ref{fig:DOS}(a) and (c). For smaller energies, the DOS of the DM approaches the power law $\propto E^{-0.5}$ of the ordered system, see Eq.~\eqref{eq:DOS}, with the agreement being almost perfect for $w=1$.  This is the first indication that the low-energy modes of the DM resemble those of the ordered system. To test this hypothesis, we evaluated the wavelength of the eigenmodes by determining their largest Fourier component. The resulting relation between wavelength and energy is shown in Fig.~\ref{fig:DOS}(b) for $w=1$ and agrees for small $E$ with that of the ordered system, with $\lambda \propto E^{-0.5}$, as in Eq.~\eqref{eq:dispersion_DM}. For $w=2$, the wavelengths are somewhat shorter than for the ordered system (Fig.~\ref{fig:DOS}(d)). 

In the RCM, the DOS of the disordered system deviates from that of the ordered system and increases for small $|E|$ , in agreement with the known singularity at $E=0$ \cite{theodorouExtendedStatesOnedemensional1976}. 
However, for $w=1$ we do not yet see the exponent -1 at low energies that was derived in \cite{theodorouExtendedStatesOnedemensional1976}. A numerical evaluation of the density of states for much smaller energies would require system sizes that are too large for the diagonalisation algorithm. The relation between wavelength and energy deviates also from the ordered system, with the deviation increasing when energy becomes small. We ascribe this to the random-walk nature of the wave functions in the presence of disorder. When evaluating the dominant wavelength, we considered only one sublattice and omitted the wavelength 4 in order to see the larger-scale amplitude modulation. The wavelength 4 captures the trivial features of the eigenfunctions, namely the decoupling of the two sublattices at $E=0$ and the change of sign between neighbouring sites of a sublattice.

\subsection{Participation ratio}
\subsubsection{Probability distribution function}
\begin{figure}[H]
     \centering
     \begin{subfigure}[b]{0.49\textwidth}
         \centering
         \includegraphics[width=\textwidth]{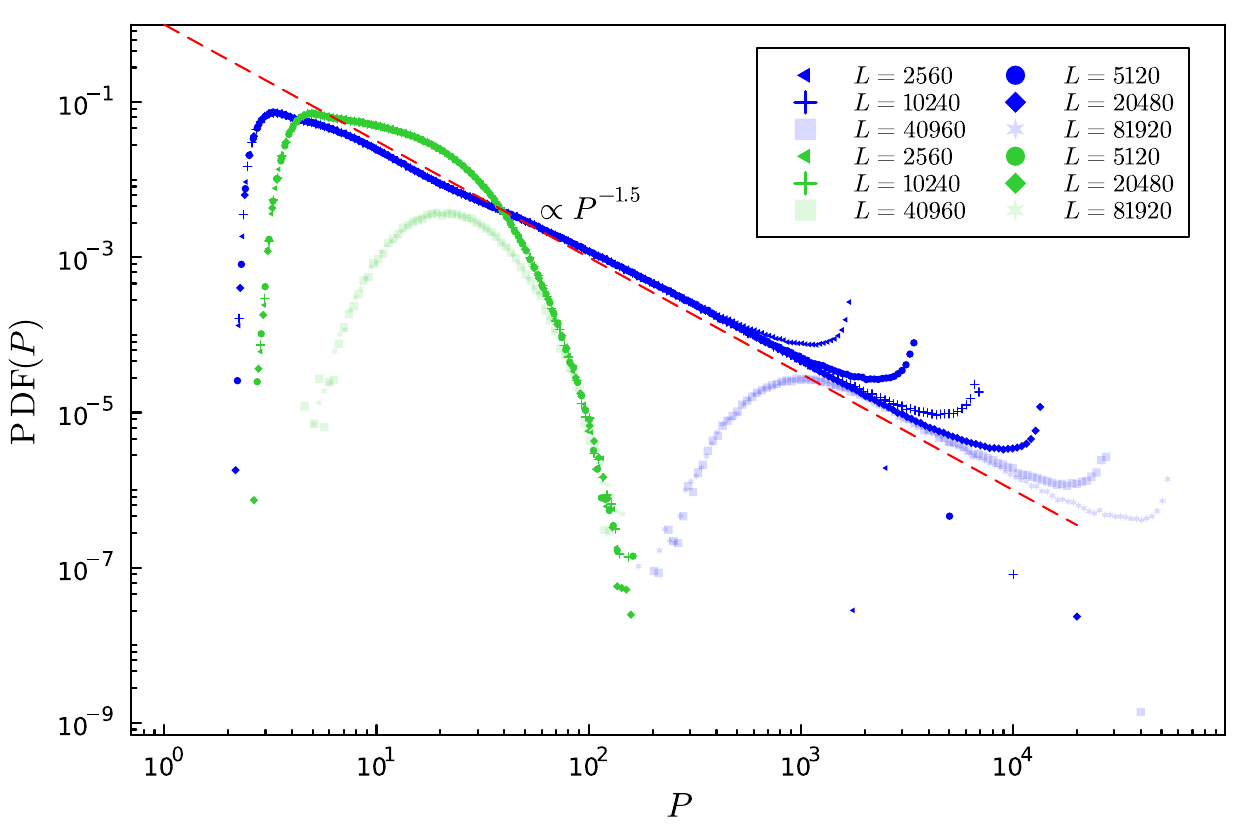}
         \caption{}
     \end{subfigure}
     \hfill
     \begin{subfigure}[b]{0.49\textwidth}
         \centering
         \includegraphics[width=\textwidth]{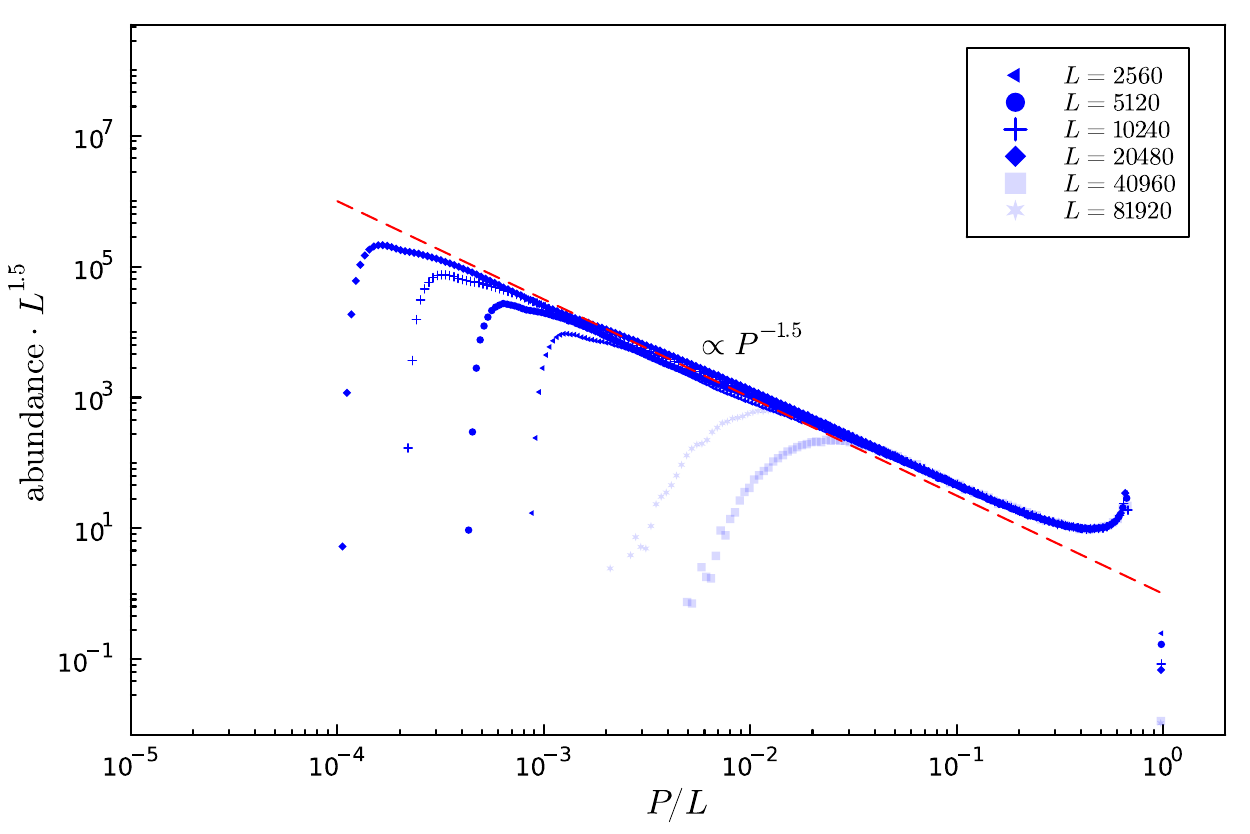}
         \caption{}
     \end{subfigure}
     \begin{subfigure}[b]{0.49\textwidth}
         \centering
         \includegraphics[width=\textwidth]{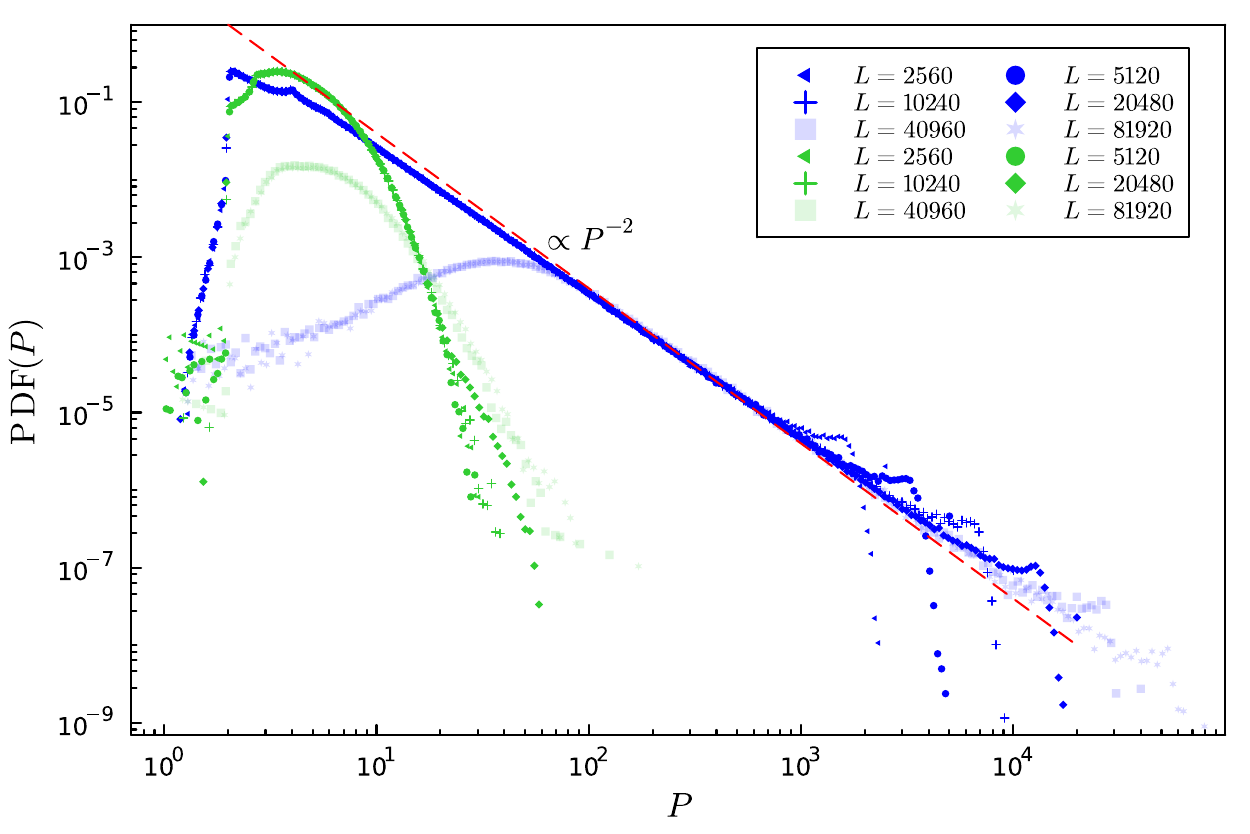}
         \caption{}
     \end{subfigure}
     \hfill
     \begin{subfigure}[b]{0.49\textwidth}
         \centering
         \includegraphics[width=\textwidth]{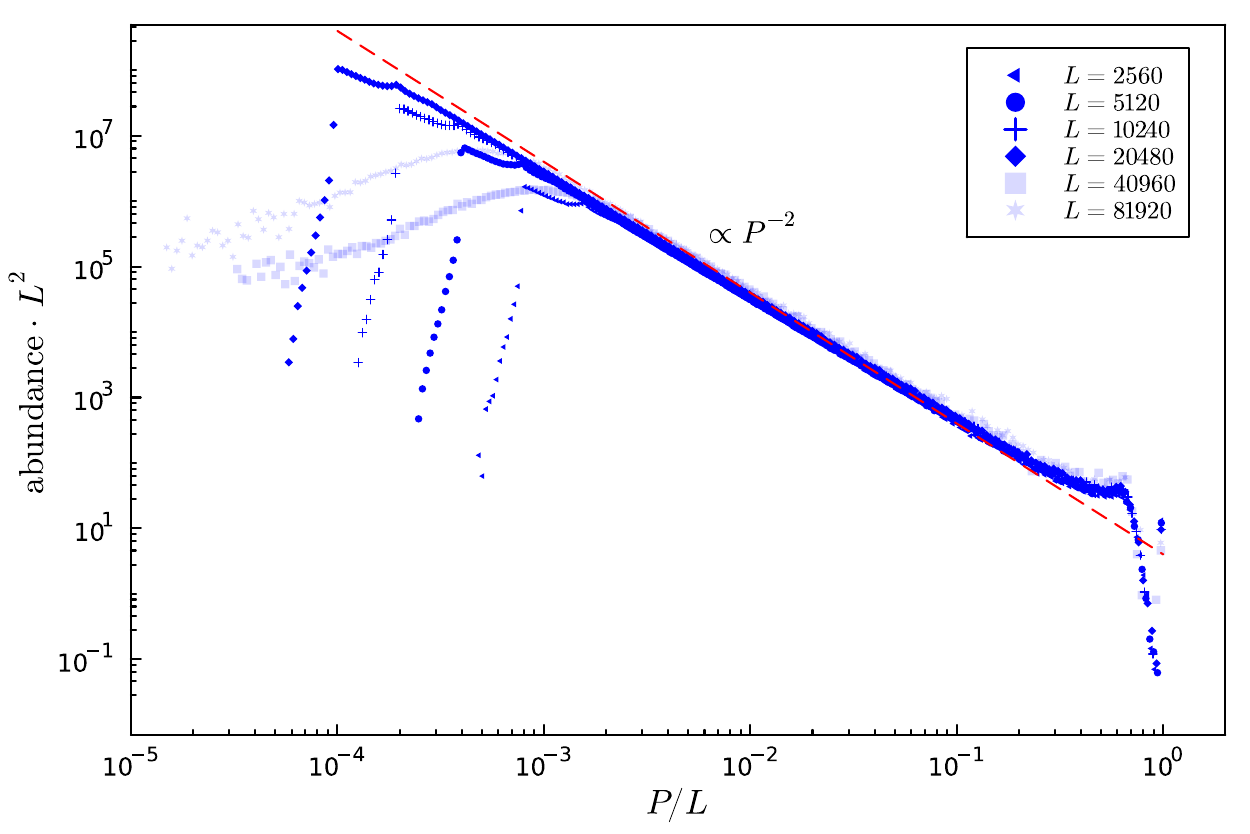}
         \caption{}
     \end{subfigure}
        \caption{Left: Probability density function of the participation ratio of the DM (blue) and RCM (green) for different system sizes $L$ for $w=1$ (a) and $w=2$ (c). The red dashed lines represent power laws with the exponent $k=-1.5$ and $k=-2$. Right: The PDFs of the DM for $w=1$ (b) and $w=2$ (d) are scaled such that the curves coincide for large $P$. The curves plotted in light blue and light green are those obtained with partial diagonalisation, where only the eigenmodes with the smallest energies were evaluated.  
        }
        \label{fig:PPDF_HC}
\end{figure}

The participation ratio $P$ is a measure of the extension of a state. When all states are localised, the distribution of the participation ratios does not depend on the system size $L$ if $L$ is large enough. This is what we observe for the RCM, see Fig.~\ref{fig:PPDF_HC}(a) and (c) (green curves). In contrast, for the DM the distribution of the participation ratio
extends over the entire system size (blue curves), which means that the DM has extended eigenfunctions that span the entire system. For small participation ratios, the curves of the DM coincide for different system sizes, indicating localised modes that are independent of the system size. For larger values of $P$, the distribution function exhibits a power law with an exponent in the vicinity of -1.5 for $w=1$ and -2 for $w=2$. Such power laws indicate a similarity of modes with different extension. The cutoff of the $P$ values scales with $L$, indicating again system-spanning modes. When the distribution is plotted as function of $P/L$ (with the appropriate rescaling of the y axis), the curves for different $L$ coincide for the larger $P$ values (Fig.~\ref{fig:PPDF_HC}(b) and (d)). In particular, the cutoff parts of the curves coincide, suggesting that the system-spanning eigenmodes for one system can be mapped onto those for a system with a different size when length is measured in units of $L$. For the largest system sizes, we used partial diagonalisation of the Hamiltonian, obtaining only the modes with the smallest energies (curves in light blue and light green). Therefore, modes with larger energies, which tend to have smaller values of $P$, are not part of those curves. 

For values $w<2$ but close to 2, there should be a crossover from the behaviour observed for $w=2$ for small system sizes (where the distribution of couplings $t_{ij}$ cannot be distinguished from that for $w=2$) to that for $w=1$ at sufficiently large system sizes. Below, when we evaluate the DM with the transfer matrix method, we will explicitly show this crossover. 

In the RCM, the distribution of $P$ values shows for $w=2$ also some particularities: There seems to be a broad scattering of the data for $P$ values close to 1, and a trend towards larger maximum $P$ values as $L$ increases. Both these features must be due the fact that  couplings $t_{n,n+1}$ can become arbitrarily small for $w=2$. This allows for large changes in the amplitude from one lattice site to the next and therefore sharper peaks. The cloud of points around $P \simeq 1$ indicates narrow wave functions of varying width. The slight increase of the weight of larger $P$ values with increasing $L$ suggests that eigenfunctions with small $E$ can have more high peaks (which make the main contribution to $P$) when the system sizes is larger. (These peaks can be seen in  Fig.~\ref{fig:eigenvectors_EM_w_2_sublattices} below.) For the largest system sizes in Fig.~\ref{fig:PPDF_HC}(b), the shape of the curves does not change anymore, and the data may have reached their true asymptotic behaviour.  

\subsubsection{Dependence on energy}
\begin{figure}[H]
     \centering
     \begin{subfigure}[b]{0.49\textwidth}
         \centering
         \includegraphics[width=\textwidth]{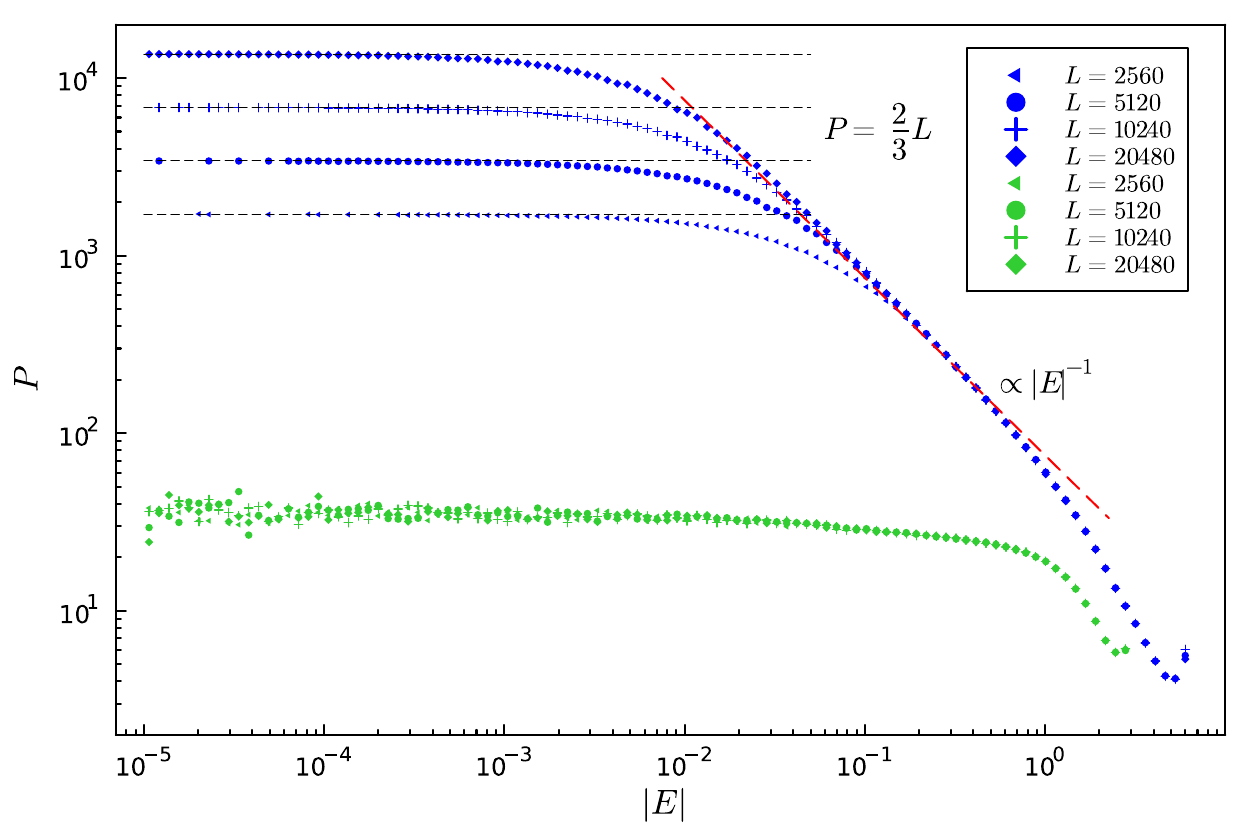}
         \caption{}
     \end{subfigure}
     \hfill
     \begin{subfigure}[b]{0.49\textwidth}
         \centering
         \includegraphics[width=\textwidth]{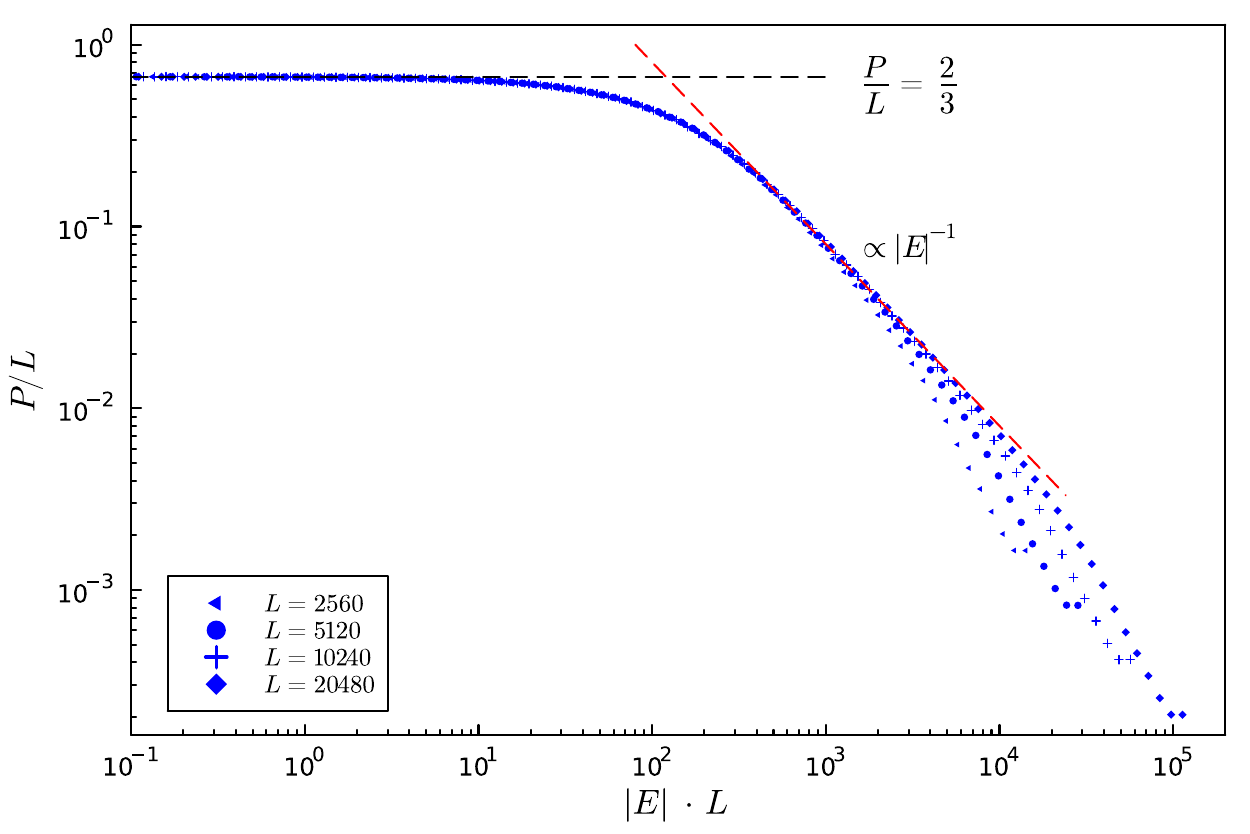}
         \caption{}
     \end{subfigure}
     \begin{subfigure}[b]{0.49\textwidth}
         \centering
         \includegraphics[width=\textwidth]{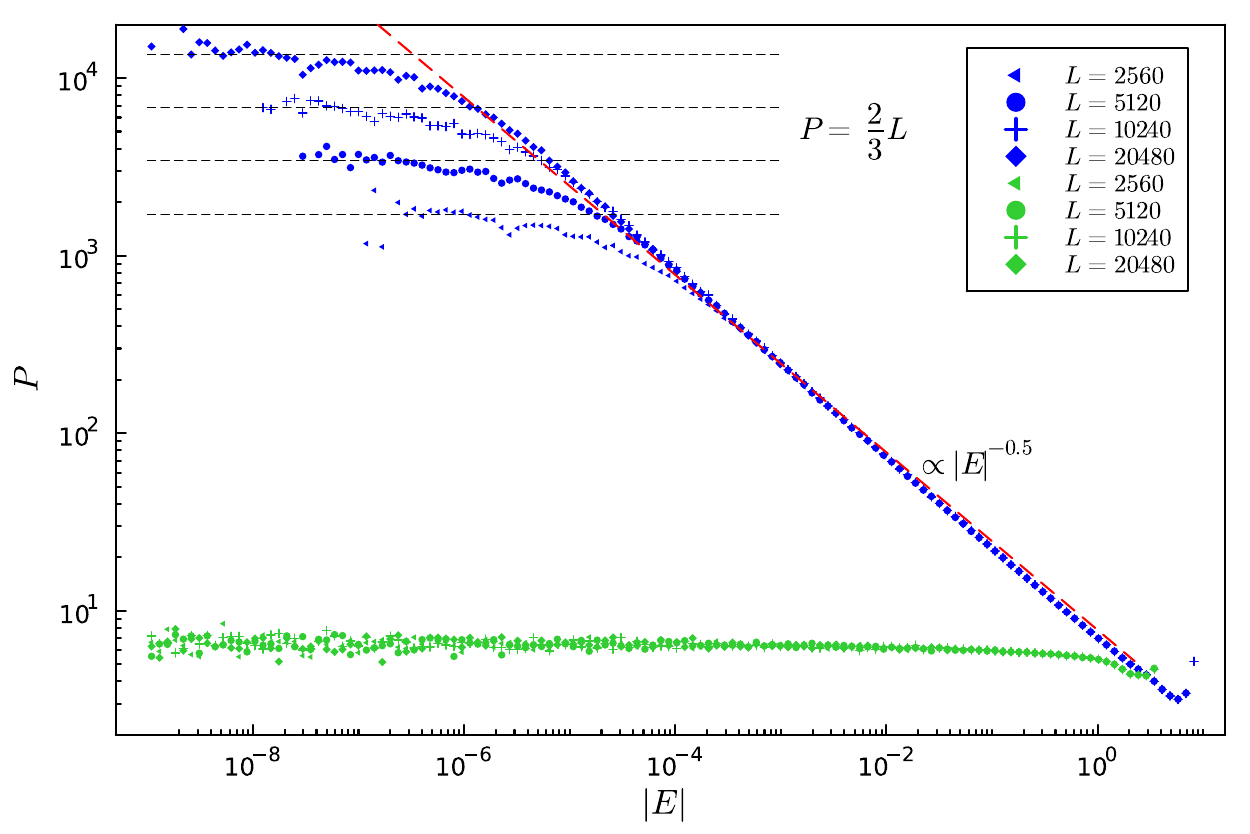}
         \caption{}
     \end{subfigure}
     \hfill
     \begin{subfigure}[b]{0.49\textwidth}
         \centering
         \includegraphics[width=\textwidth]{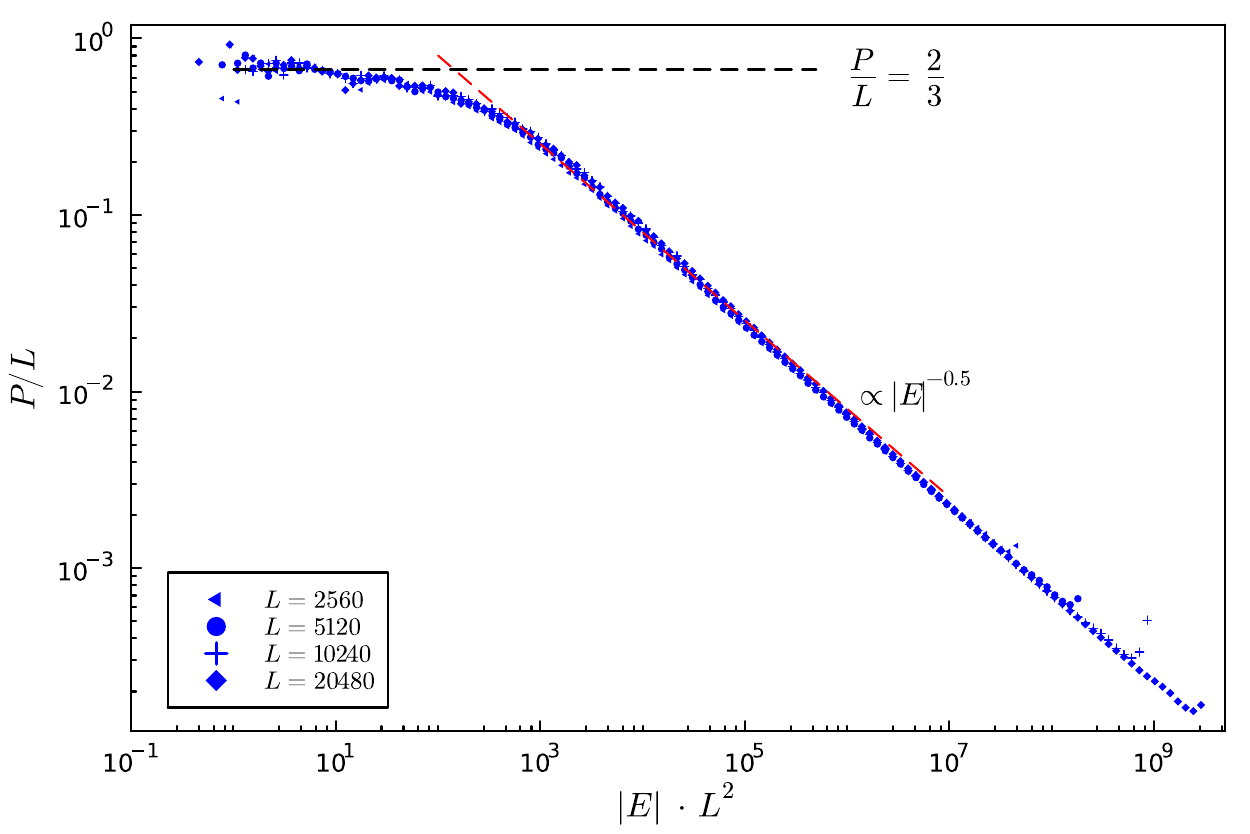}
         \caption{}
     \end{subfigure}
        \caption{Left: Dependence of the participation ratio on the energy of the DM (blue) and RCM (green) for $w=1$ (a) and $w=2$ (c) for different system lengths $L$. The dashed red lines represent power laws with exponent $k=-1$ and $k=-0.5$ respectively. Right: When $P/L$ is plotted vs $|E|\cdot L$ for $w=1$ (b) and $|E|\cdot L^2$ for $w=2$ (d) for the DM, the distributions align except for the largest energies.}
        \label{fig:EvP_HC}
\end{figure}

The dependence of the participation ratio $P$ on the energy $E$ for the DM shows again the system-spanning modes and a scaling regime, Fig.~\ref{fig:EvP_HC}(a) and (c) (blue data points). 
For small energies, $P$ approaches the constant $2/3 L$, which is the participation ratio of the eigenmodes of the ordered system, which are sine functions in 1D, i.e. extended states. For $w=2$, the data points scatter more strongly around 2/3, with a tendency to also show larger values of $P$. This is also reflected in the probability distribution function (Fig.~\ref{fig:PPDF_HC}): For $w=1$, there is a clear gap between $P=2/3$  
and the constant solution which has $P=1$, while for $w=2$ some data points are found within this gap.
For larger $E$, the participation ratio $P$ scales as $P\propto|E|^{-1}$ for $w=1$ and as $P\propto |E|^{-0.5}$ for $w=2$. The modes in the scaling regime are localised but span a large number of lattice sites. 
The power law $P\propto|E|^{-1}$ for $w=1$ is directly connected to the power law ${\mathrm{PDF}}(P)\propto P^{-1.5}$ through the density of states: from ${\mathrm{PDF}}(P)\mathrm{d}P={\rm DOS}(E)\mathrm{d}E$ follows ${\mathrm{PDF}}(P)={\mathrm{DOS}}(E)\mathrm{d}E/\mathrm{d}P$ and $\mathrm{d}P/\mathrm{d}E={\mathrm{DOS}}(E)/{\mathrm{PDF}}(P) \propto 1/E$. Performing the equivalent calculation for $w=2$, one obtains from ${\mathrm{PDF}}(P)\propto P^{-2}$ the relation $\mathrm{d}P/\mathrm{d}E={\mathrm{DOS}}(E)/{\mathrm{PDF}}(P) \propto 1/E^{0.5}$
For the largest energies, where the modes span only a small number of lattice sites, the power law breaks down, and the curves become steeper.
The hook at the right end of the energy band is outside the band of the ordered system (where $E_{\mathrm{max}}=2$) and is a so-called Lifshitz tail, which arises with bounded disorder when a cluster of several neighbouring sites has energy values (and in the DM also coupling values) close to the edge of the disorder distribution~\cite{lifshitz_energy_1964, johri_singular_2012, johri_singular_twosite_2012}. States localised on such clusters have a larger energy than in the ordered system, and their participation ratio is of the order of the size of the cluster. 

The curves $P(E)$ of the DM coincide for the different system sizes in the transition regime from extended to large localised states when $P/L$ is plotted versus $E\cdot L$ for $w=1$  (Fig.~\ref{fig:EvP_HC}(b)) and versus $E\cdot L^2$ for $w=2$  (Fig.~\ref{fig:EvP_HC}(d)). 
This means for $w=1$ that the largest energy of the system-spanning modes scales as $E_{\mathrm{maxspan}}\propto 1/L$. 
Due to the dispersion relation $E\propto k^2$ for small $E$, this implies that the largest $k$ value for system-spanning modes scales as $1/\sqrt{L}$. The $k$ values are equally spaced and have values $\frac {2\pi n}L$ with integer $n=1,\dots,L$.  Therefore, the total number of system-spanning modes scales as $\sqrt{L}$, in agreement with the results of analytical calculations reported in the literature \cite{dunlap_absence_1989}. For $w=2$, the same type of consideration leads to the result that the total number of system-spanning modes scales as $L^0$, that is, it remains finite in the limit of infinite system size or at least increases slower than any power law. This is the reason why the power law and the scaling collapse in Fig.~\ref{fig:EvP_HC}(d) extend up to the largest energy values. For $w=1$, the $\sqrt L$ largest energy values belong to system-spanning modes and are not part of the scaling regime.

For the RCM, the relation between $P$ and $E$ is independent of $L$, see Fig.~\ref{fig:EvP_HC}(a) (green data points), which is expected if all modes are localised. 
The data points for $P$ show considerable scattering when $|E|$ is smaller than $\approx 10^{-2}$, with the mean approaching a constant with decreasing $|E|$. This constant is smaller for $w=2$ than for $w=1$. 

\subsection{Features of the eigenmodes}
\begin{figure}[H]
     \centering
     \includegraphics[width=\textwidth]{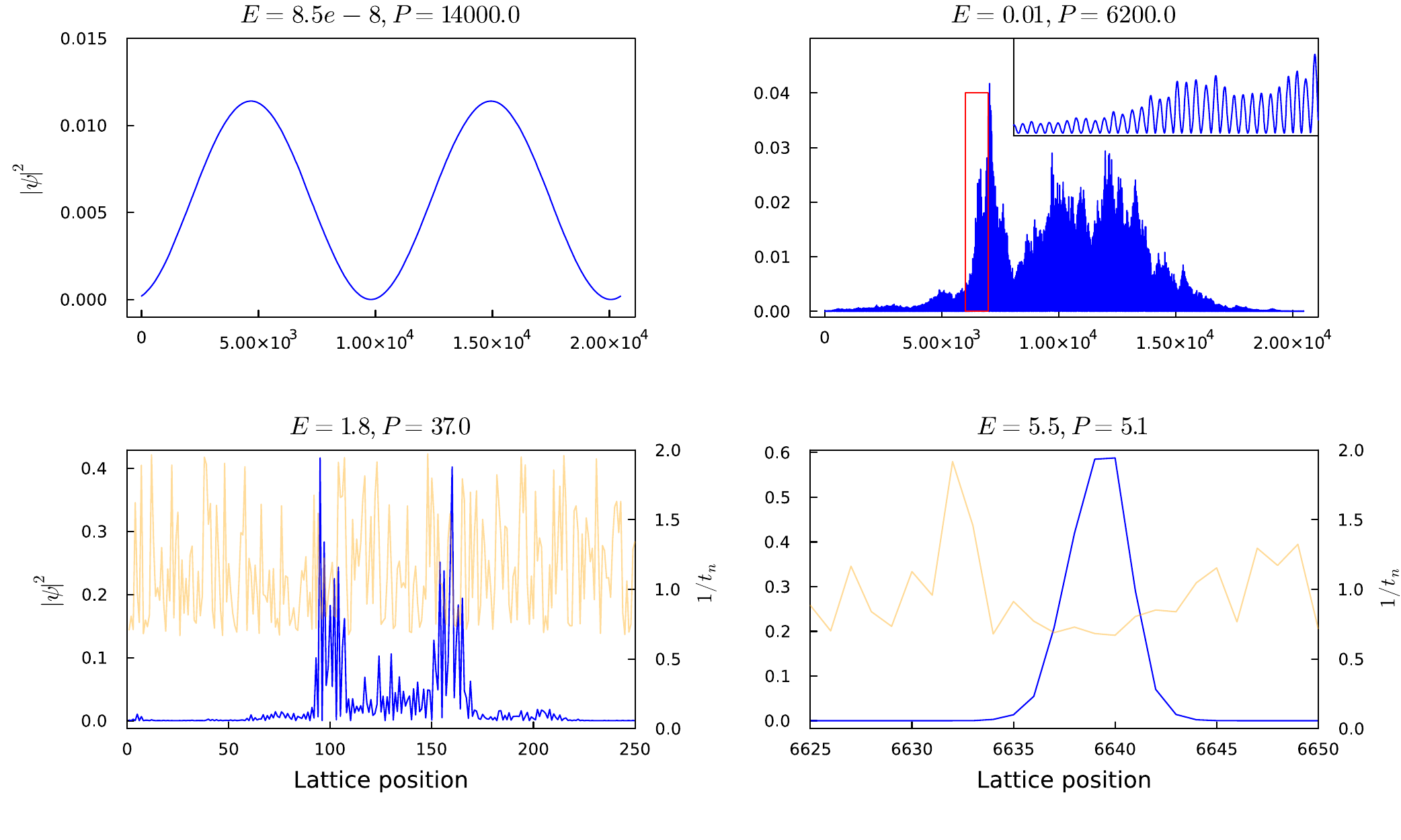}
    \caption{Eigenvectors in different localisation regimes of the DM with $L=20480,~w=1$. The light brown data in the last two graphs show additionally the inverse couplings $1/t_{n,n+1}$ (secondary $y$-axis).
        }
    \label{fig:eigenvectors_LM_w_1}
\end{figure}

A look at the eigenmodes confirms and deepens the understanding gained from the data for the participation ratios. For the DM, there are system-spanning eigenmodes that resemble sine functions, see Fig.~\ref{fig:eigenvectors_LM_w_1}(a). The randomness of the model is barely visible in these functions when $E$ is very small and $w=1$, but becomes more visible for larger $E$, see Fig.~\ref{fig:eigenvectors_LM_w_1}(b). 
The eigenfunction in Fig.~\ref{fig:eigenvectors_LM_w_1}(c) is localised but covers a large number of lattice sites. This is a mode that belongs to the scaling regime.  The last shown eigenfunction is the one with the largest energy; it has a single peak.

\begin{figure}[H]
     \centering
     \includegraphics[width=\textwidth]{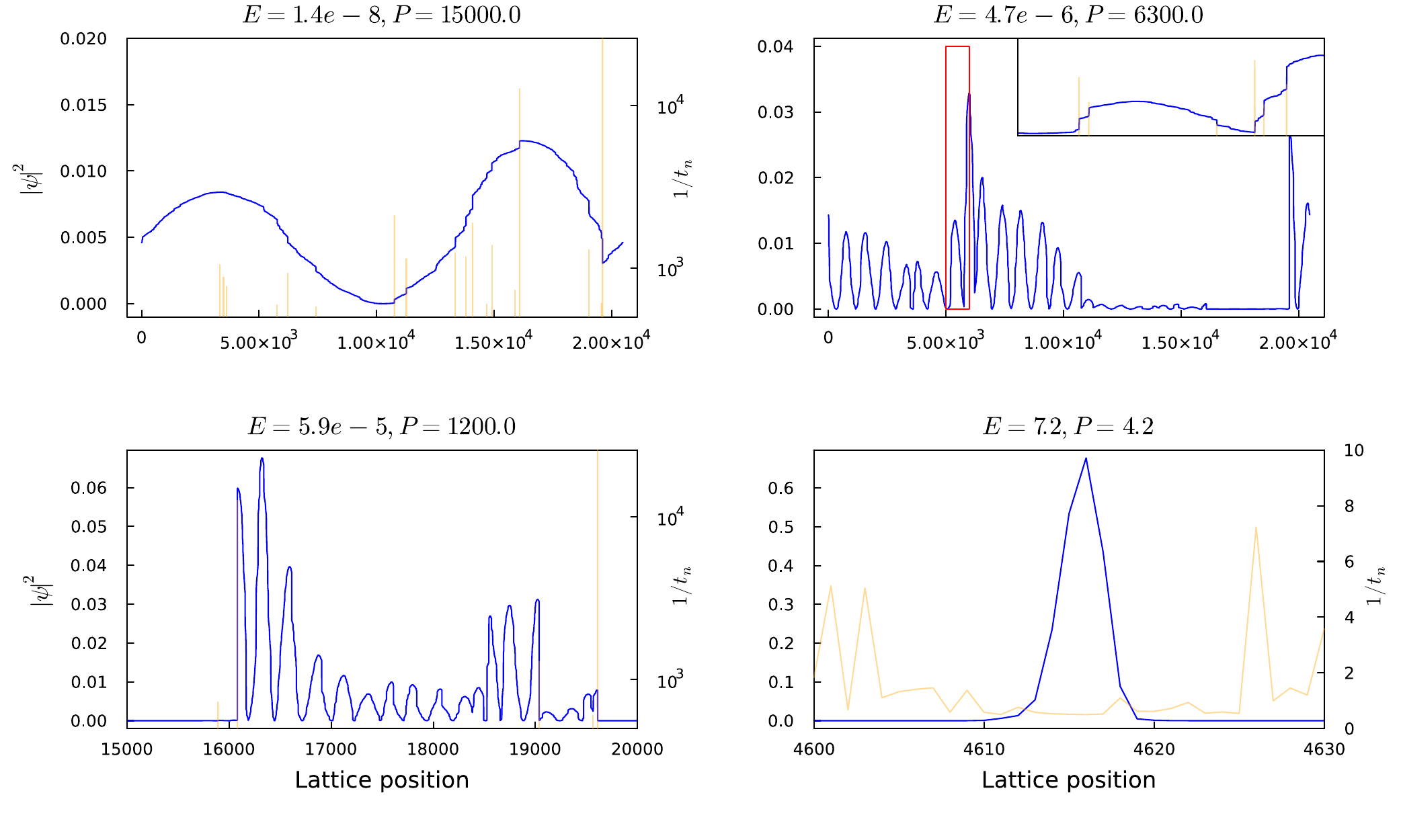}
    \caption{Eigenvectors in different localisation regimes of the DM with $L=20480,~w=2$. The light brown data additionally shows the inverse couplings $1/t_{n,n+1}$ (secondary $y$-axis).
    }
    \label{fig:eigenvectors_LM_w_2}
\end{figure}

For $w=2$, the eigenmodes of the DM show the same trends as for $w=1$ with increasing $E$ (Fig.~\ref{fig:eigenvectors_LM_w_2}), but due to the wide range of inverse coupling constants $1/t_{n,n+1}$, the eigenfunctions have larger jumps. For the system-spanning mode, one can see that this makes the wavelength smaller compared to what it would be with less disorder. The deviations from the shape of the sine function caused by the large jumps are also responsible for the scattering of the $P$ values of the system-spanning modes seen in the previous figures for $w=2$.  

\begin{figure}[H]
     \centering
     \includegraphics[width=\textwidth]{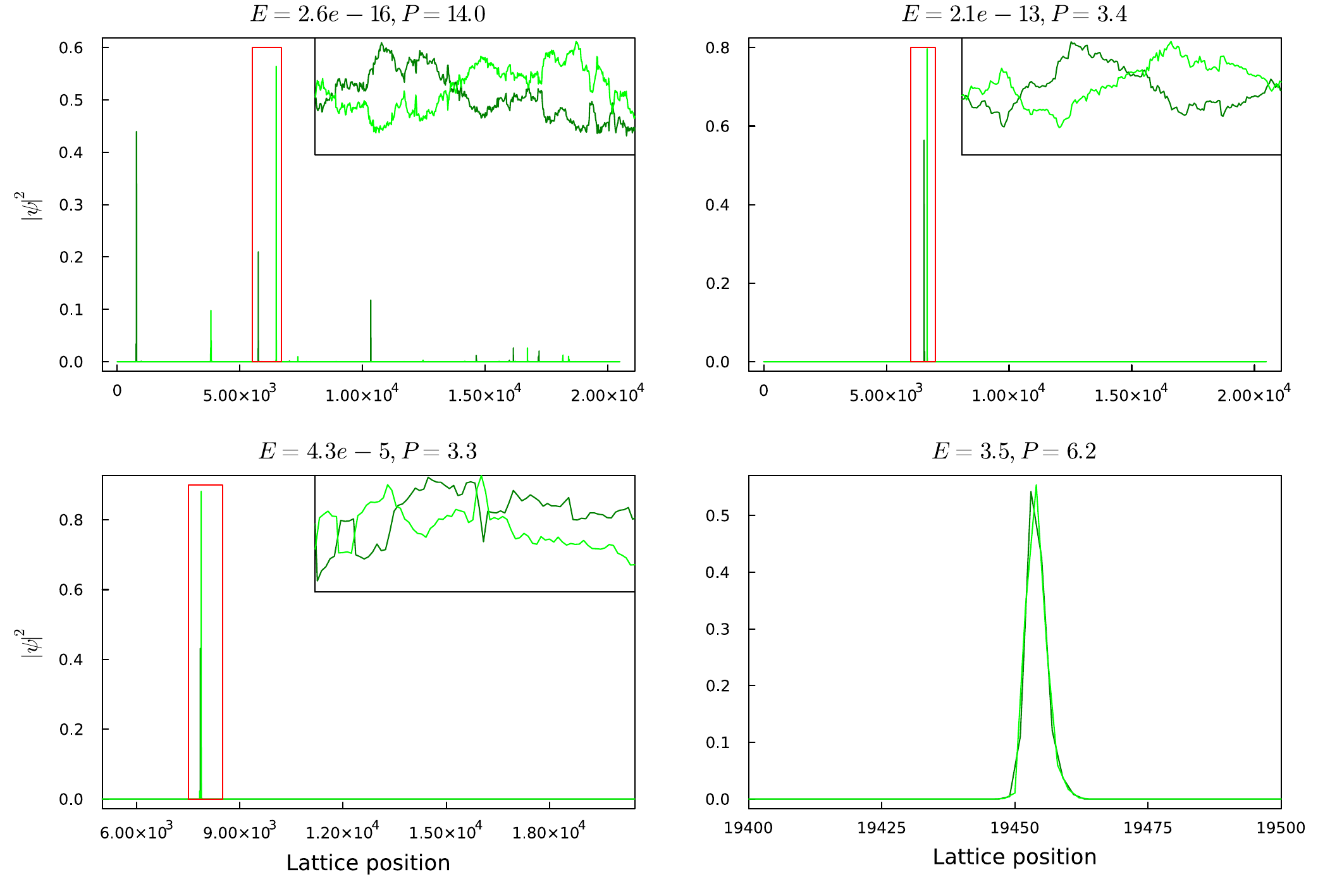}
    \caption{Eigenvectors in different localisation regimes of the RCM with $L=20480,~w=2$. The two colours indicate the two sublattices. 
    The insets show the segment in the red rectangles and have a logarithmic y-axis. 
    }
    \label{fig:eigenvectors_EM_w_2_sublattices}
\end{figure}

For the RCM, all eigenmodes are localised, but extend over more lattice sites for smaller $|E|$, see Fig.~\ref{fig:eigenvectors_EM_w_2_sublattices}.  We show only eigenmodes for $w=2$ since there is no striking difference between $w=2$ and $w=1$. For smaller $|E|$, the contributions of the two sublattices to the eigenmode appear to become decoupled, with two different main peaks, with a distance that increases with decreasing $|E|$. We can explain this with the fact that  $\ln |\psi_n|$ on the two sublattices perform each a random walk for $E=0$, see Eq.~\eqref{eq:independent}.  The eigenfunctions for small $|E|$ are therefore very close to random walks. Because the eigenfunctions must be normalised, these random walks must return to the origin. Even though the mean length of  returning random walks diverges (see e.g.~\cite{ibeRandomWalk2013}), the main weight of $|\psi_n|^2$ is confined to those lattice sites that belong to the largest deviation, giving rise to the apparently sharp peak in the wave function, and these are the lattice sites that make the main contribution to the participation ratio.    
  The lengths and maximum deviations of returning random walks have a broad distribution, and this is reflected in the scattering of the values of $P$ for small $|E|$.

\subsection{Localisation length}
\begin{figure}[H]
     \centering
     \begin{subfigure}[b]{0.49\textwidth}
         \centering         \includegraphics*[width=\textwidth]{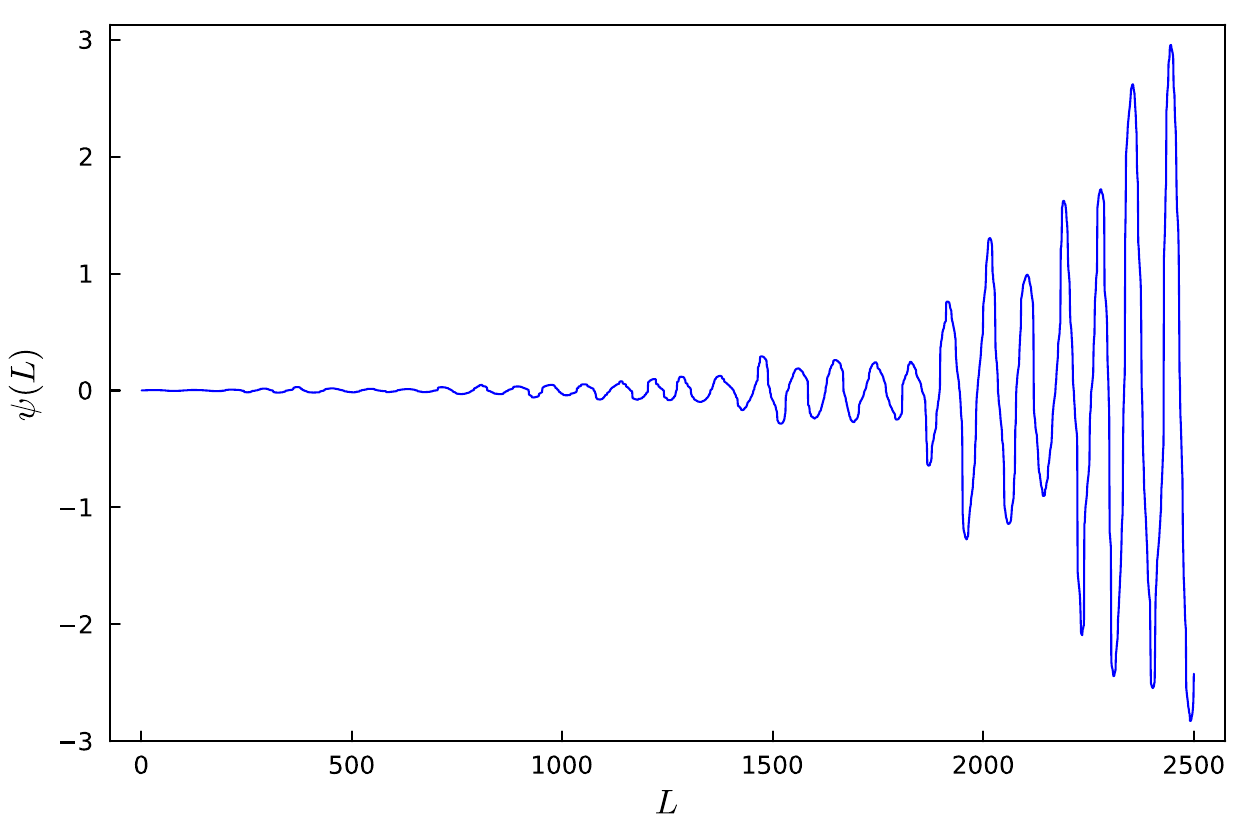}
         \caption{}
     \end{subfigure}
     \hfill
     \begin{subfigure}[b]{0.49\textwidth}
         \centering
         \includegraphics*[width=\textwidth]
         {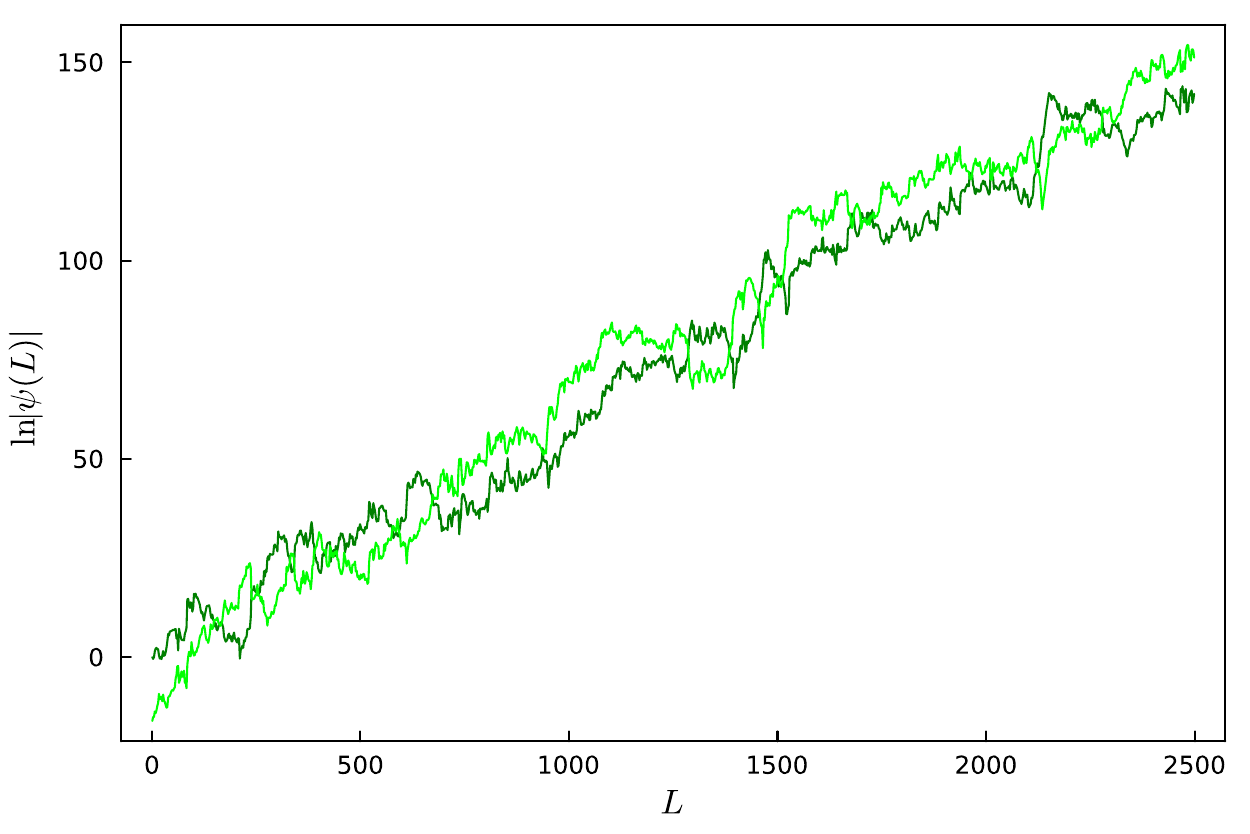}
         \caption{}
     \end{subfigure}
        \caption{(a) Example of $\psi(L)$ (divided by $\num{e4}$) resulting from the transfer matrix calculation for the DM, for the parameters $w=1.95$ and $E=0.0025$, showing the periodic oscillation and the exponential increase of the wave function. (b) Example of $\psi(L)$ for the RCM, for the parameters $w=2$ and $E=10^{-7}$. The two colours indicate even and odd lattice sites. 
        }
        \label{fig:TMM_examples}
\end{figure}

Examples of wave functions resulting from the transfer matrix method are shown in Fig.~\ref{fig:TMM_examples}, illustrating for the DM the periodic oscillations combined with an exponential increase of the amplitude, and for the RCM the exponential increase with only weakly coupled sublattices. In both cases, the exponential increase holds on average for a large sample or for a sufficiently large number of iterations. Individual runs show fluctuations around this exponential increase. 

\begin{figure}[H]
     \centering
     \begin{subfigure}[b]{0.49\textwidth}
         \centering
         \includegraphics[width=\textwidth]{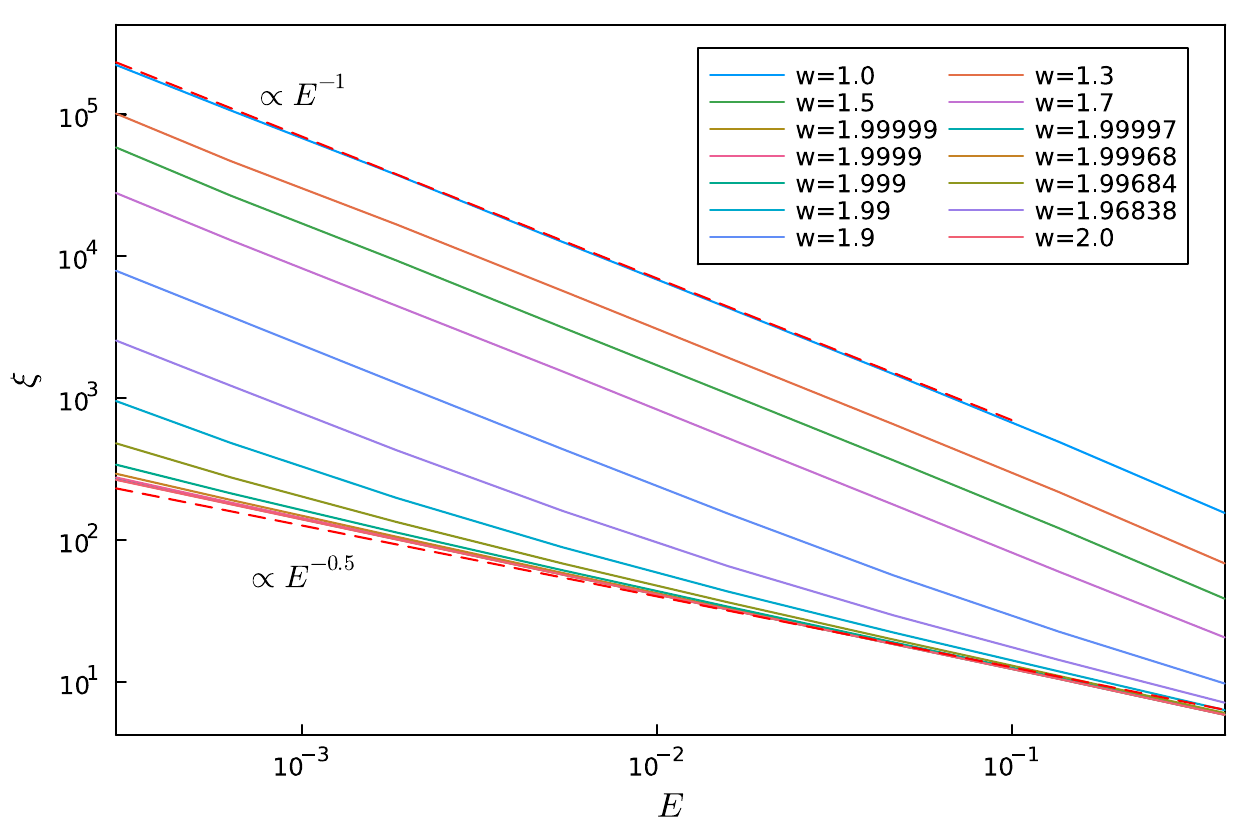}
         \caption{}
     \end{subfigure}
     \hfill
     \begin{subfigure}[b]{0.49\textwidth}
         \centering
         \includegraphics[width=\textwidth]{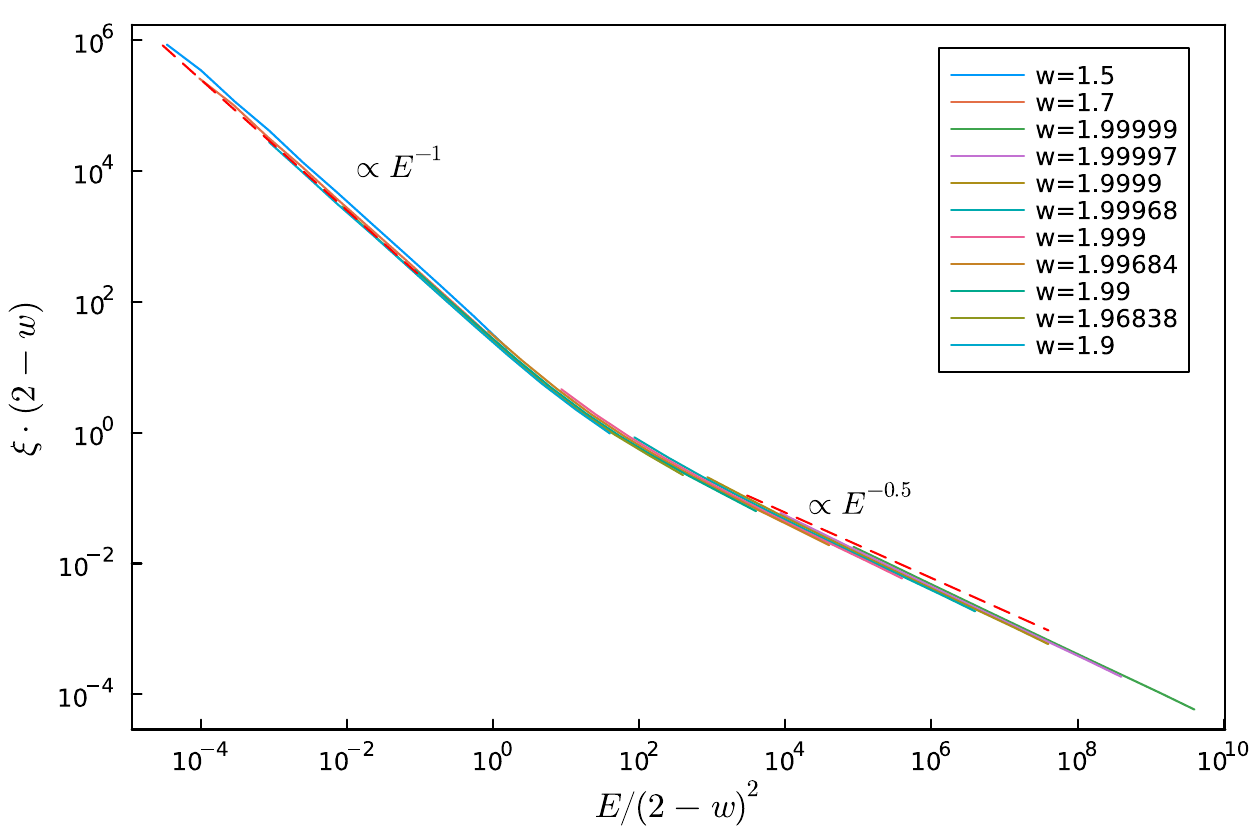}
         \caption{}
     \end{subfigure}
        \caption{(a) Dependence of the localisation length on the energy in the DM for different disorder strengths $w$. (b) The data for $w$ close to 2 collapse onto one curve when $E$ is plotted in units of $(2-w)^{2}$ and $\xi$ in units of $(2-w)^{-1}$.      
        }       
        \label{fig:XiE_DM_different_w}
\end{figure}

We evaluated the localisation length $\xi$ as a function of $E$ based on Eq.~\eqref{defxi}, choosing a sufficiently large number $L$ of iterations of the TMM to be in the asymptotic regime, where finite-size effects and stochastic fluctuations are no longer perceived. When the strength of the disorder $w$ is close to 2, the localisation length of the DM shows a crossover between a slope of $-1$ and a slope of approximately $-0.5$ as $E$ increases, see Fig.~\ref{fig:XiE_DM_different_w}. For $w=2$, the slope is close to $-0.5$, for $w=1$, it is $-1$. This means that $\xi$ depends in the same way on $E$ as the participation ratio $P$ shown above in Fig.~\ref{fig:EvP_HC}. The data for different $w$ (if close to 2) fall on one master curve when $\xi$ is measured in units of $(2-w)^{-1}$ and $E$ in units of $(2-w)^{2}$. Since the energy is inversely proportional to the square of the wavelength $\lambda$ (see the dispersion relation in Fig.~\ref{fig:DOS}(b) and (d) above), measuring $E$ in units of $(2-w)^{2}$ means measuring $\lambda$ in units of $(2-w)^{-1}$, as is done for $\xi$. Thus, there is a characteristic length scale of $(2-w)^{-1}$ in the system. This is the length beyond which the lower cutoff due to $w<2$ of the coupling values $t_{n,n+1}$ becomes visible. Below this length, the distribution of the coupling values is like that of $w=2$, since the typical smallest coupling for $w=2$ is of the order of the inverse size of the sample from which the coupling values are chosen.  

\begin{figure}[H]
     \centering
     \begin{subfigure}[b]{0.49\textwidth}
         \centering
         \includegraphics[width=\textwidth]{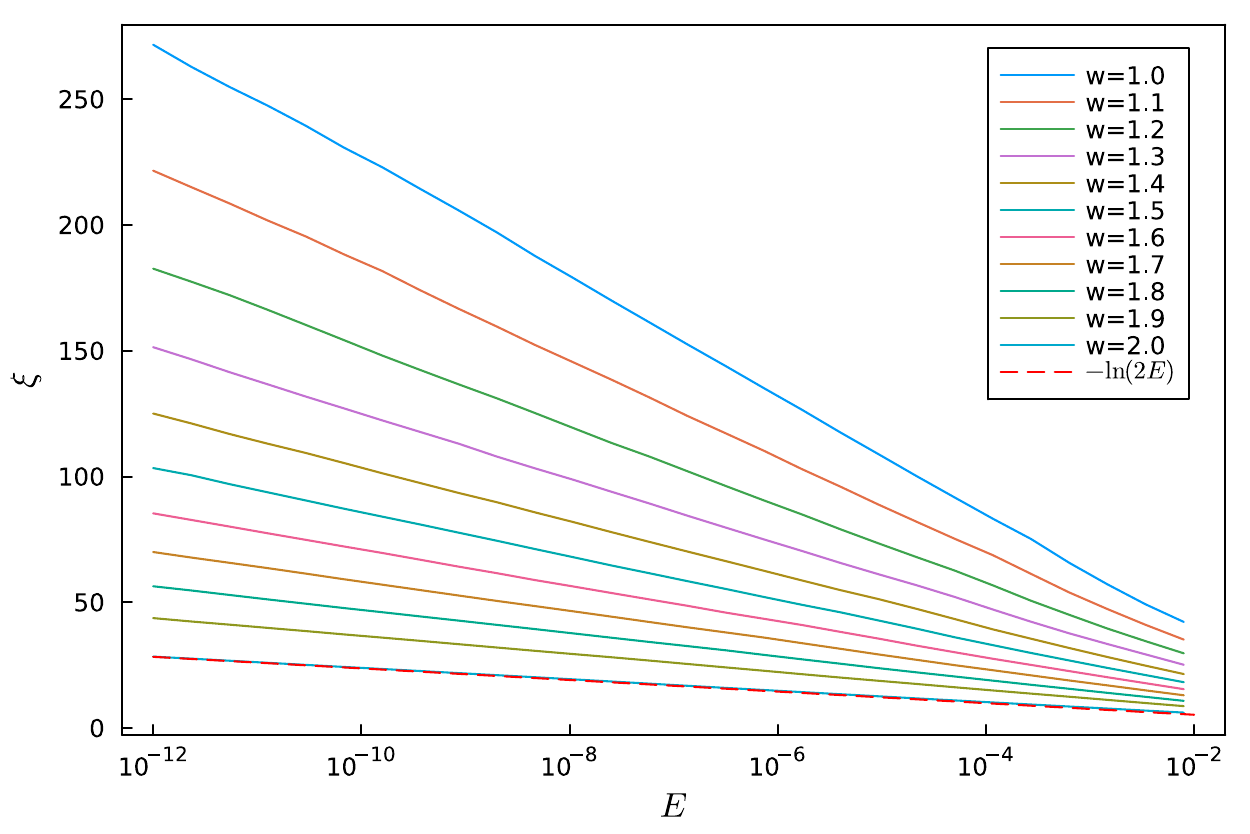}
         \caption{}
     \end{subfigure}
     \hfill
     \begin{subfigure}[b]{0.49\textwidth}
         \centering
         \includegraphics[width=\textwidth]{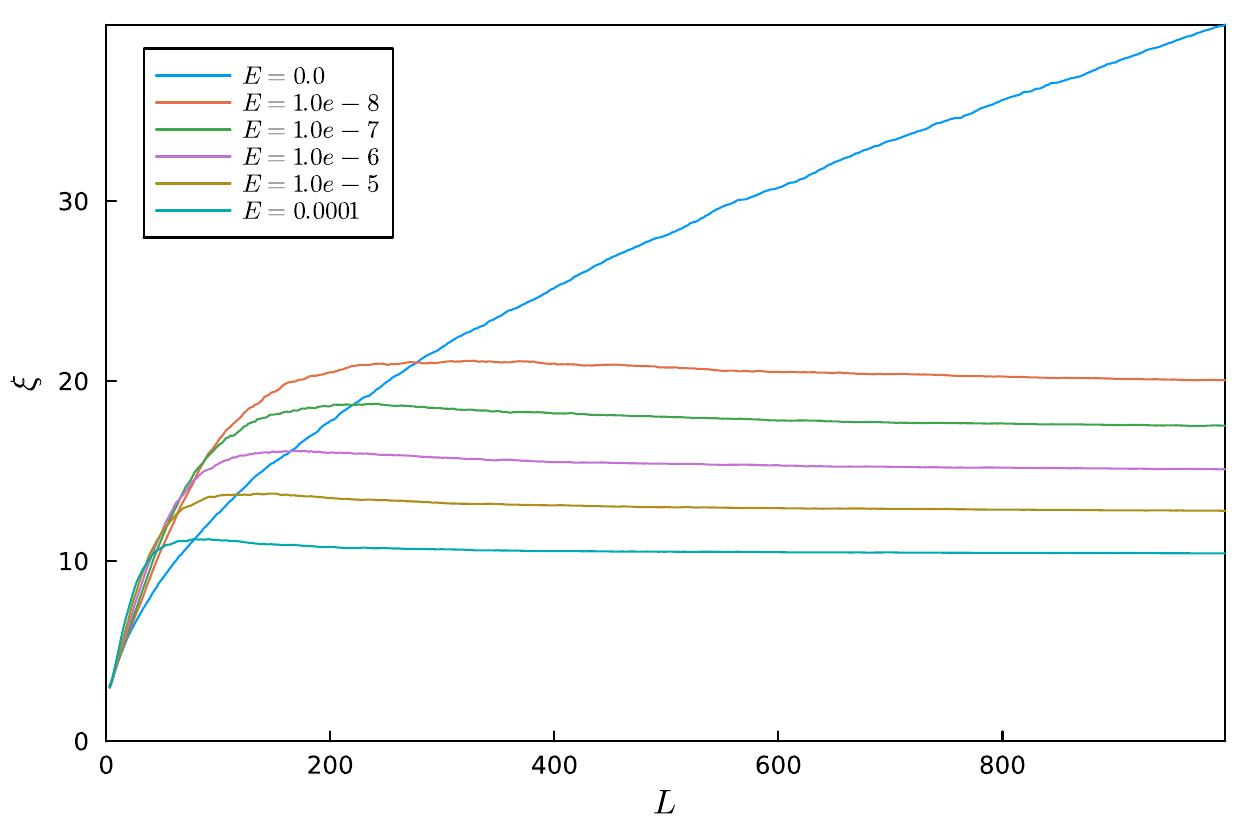}
         \caption{}
     \end{subfigure}
        \caption{(a) Dependence of the localisation length on the energy in the RCM for different $w$ (logarithmic $x$-axis). The red dashed line is a fit for $w=2$.
        (b) Localisation length versus $L$ for the RCM for $w=2$ and different $E$. It was obtained by averaging the absolute value of $\ln|\psi|$ over 10,000 runs, and dividing $L$ by the result.}
        \label{fig:Xi_RCM}
\end{figure}

 For the RCM, the localisation length is small for the chosen values of $w$ and depends logarithmically on $1/E$, i.e., $\xi = a \log{E}$ with the factor $a$ depending on $w$, see Fig.~\ref{fig:Xi_RCM}(a). For $E=0$, the logarithm of the wave function performs a random walk, as derived above. Its ensemble mean is zero, but the average deviation from the starting value increases as $\sqrt{L}$. This means that the localisation length diverges for $E=0$ as $\sqrt L$, as shown in Fig.~\ref{fig:Xi_RCM}(b). For small nonzero $E$, the data follow for small $L$ those for $E=0$, and for larger $L$ they  approach a constant value.

\section{Theoretical considerations and scaling arguments}
The two models investigated in this paper differ fundamentally in their properties. While the DM shows of the order of $\sqrt{L}$ system-spanning modes for $w<2$, all modes in the RCM are localised, with the localisation length obtained from the calculation of the transfer matrix diverging logarithmically as $E \to 0$. In contrast, the localisation length obtained for the Anderson model remains finite in the limit $E \to 0$ when the strength of the disorder is finite~\cite{kramer_localization_1993}. 

In the following, we will provide qualitative explanations for the various results obtained so far, and the relations between them.

\subsection{System-spanning modes in the DM}

The existence of system-spanning modes in the DM can be explained by the conservation law of this model. In the version Eq.~\eqref{DMt} of this model, the total amount $\sum_n c_n$ of the diffusing substance is conserved. If the system is initially perturbed from equilibrium by moving part of the substance from the left half of the system to the right half, equilibrium can only be restored by a transport of substance over a distance of the order of (half) the system size. This means that there must be system-spanning relaxing modes in this model. We have seen above that for $w<2$ there are $\propto \sqrt{L}$
system-spanning modes, with the largest energy of system-spanning modes being $E_{\mathrm{maxspan}} \propto 1/L$. 

\subsection{Scaling of the participation ratio with the energy in the DM}

Those modes that do not span the system but extend over a large number of sites show a scaling behaviour $P \propto 1/E \propto \lambda^2$ for $w<2$, when the couplings $t_{n,n+1}$ between neighbours cannot be arbitrarily small (Fig.~\ref{fig:EvP_HC}). In the following, we give an intuitive argument for this result: In contrast to the ordered system, where all modes are system-spanning and have a constant amplitude, the disordered system shows fluctuations in the amplitude due to the broad distribution of coupling constants. An eigenmode of equation \eqref{DMt} relaxes according to $e^{-t/\tau}$ but does not change its shape. In the ordered system, the proportion of substance that flows per unit time from a mound of the eigenmode to the two neighbouring valleys is the same for both valleys and the same everywhere along the eigenmode. In the disordered system, the proportion of substance that flows per unit time from mounds to valleys differs between mounds. 
The average time that the flowing substance needs to flow from site $n$ to site $n+1$ is proportional to $1/t_{n,n+1}$. The time to flow from $n$ to $n+l$ is then $\sum_{m=n}^{n+l} 1/t_{m,m+1}$, and the effective coupling constant for this distance is the inverse of this time. 
The flow rate from a mound to a valley is determined by the height difference between the mound and the valley, multiplied by the effective coupling constant over the distance $\lambda/2$, which is $1/(\sum_{n=1}^{\lambda/2} 1/t_{n, n+1})$. This effective coupling has a standard deviation $\propto 1/\sqrt{\lambda}$ due to the statistical independence of the bond strengths $t_{n,n+1}$. The strengths of neighbouring effective couplings thus differ by a factor of the order $(1\pm 1/\sqrt{\lambda})$. Since the shape of a relaxing eigenmode does not change with time, its amplitude must vary such that the proportion of substance flowing out of each mound per unit time is the same for each mound, i.e., the amplitude change from a wave mound to the next valley must also be of the order of $(1\pm 1/\sqrt{\lambda})$. The amplitude thus performs a random walk, and it takes of the order of $\lambda$ wavelengths to go from the amplitude of the highest mound to an amplitude zero at the end of the eigenmode. This means that the eigenmode has an extension $\propto \lambda^2$.  

This argument does not hold for $w=2$, as arbitrarily small couplings may occur that block the flow of substance. The inverse of the effective coupling constant over a distance $l$ increases logarithmically with $l$ instead of being Gaussian distributed around a finite mean. Very small couplings that act as flow barriers are natural boundaries of relaxing modes, as can be seen in Fig.~\ref{fig:eigenvectors_LM_w_2}. Below, when discussing the TMM, we will argue that this reduces the localisation length to the order of the wavelength.

\subsection{Scaling of the localisation length with energy in the DM}
The recursion relation Eq.~\eqref{eqTM} for the DM can be written as
\begin{equation}
    \psi_2-\psi_1 = \frac{t_{01}}{t_{12}} (\psi_1-\psi_0)- \frac {E} {t_{12}}\psi_1~.
\end{equation}
Iteration of this relation gives
\begin{equation}
    \psi_L - \psi_{L-1} = (\psi_1-\psi_0)\frac{t_{0,1}}{t_{L-1,L}} - \frac {E} {t_{L-1,L}}\sum_{l=1}^{L-1}\psi_l \,
\end{equation}
which can for small $E$, i.e. large wavelengths, be approximated by the continuum expression
\begin{equation}
    \psi_L' = \psi_1'\frac{t_{01}}{t_{L-1,L}}  - \frac E {t_{L-1,L}}\int_{L=1}^{L-1}\psi_l  \mathrm{d}l\, .\label{recursionDMcont}
\end{equation}
In the absence of disorder, and by taking the derivative on both sides, we obtain $\psi_L'' = -E\psi_L$, which has the known solution $\psi_L=\sin(\sqrt{E}L)$. The integral in the recursion relation Eq.~\eqref{recursionDMcont} becomes zero whenever $L$ is a multiple of the wavelength $\lambda$ in the absence of disorder. When disorder is included, the amplitude of the periodic oscillation changes, and the integral does not vanish anymore when $L$ is a multiple of $\lambda$. We therefore search for a different solution. This solution is an oscillating function with an exponentially increasing amplitude; see Fig.~\ref{fig:TMM_examples}(a).

To obtain this solution, we need a relation between the (positive) slope $\psi'_{n\lambda}$ at the zero of the wave function and the (negative) area $S_n \equiv \int_{n\lambda}^{(n+1)\lambda} \psi_l \mathrm{d}l$ under the next section of the wave: It is $S_n = -a\cdot \lambda \cdot \psi'_{n\lambda}$ with some constant $a$ since the area is proportional to the wavelength and to the initial slope. 
Let $N\coloneqq \frac{L}{\lambda}$ be the number of wavelengths that fit into the system length.
We thus obtain from \eqref{recursionDMcont}
\begin{equation}
    S_N \simeq \frac a \lambda \sum_{n=1}^{N-1} S_n
\end{equation}
with the solution
\begin{equation}
    S_N \simeq e^{\frac a \lambda N} = e^{\frac a \lambda \frac L \lambda} \equiv e^{\frac {L}{\xi}}\, . 
\end{equation}
The localisation length $\xi$ increases with decreasing $E$ as $\xi \propto \lambda^2 \propto 1/E$, in the same way as the participation ratio. 

Our data above (Fig.~\ref{fig:XiE_DM_different_w}) reveal that this is correct only for $w<2$ and for sufficiently small $E$ such that the wavelength is larger than $1/(2-w)$. For $w=2$, the coupling strengths $t_{l,l+1}$ can become arbitrarily small. Over one wavelength, the largest term $1/t_{l,l+1}$ is of the order of $\lambda$, which is very large for very small $E$. If we write the recursion relation Eq.~\eqref{eqTM} in the form
\begin{equation}
 \psi_l=\psi_{l-1} + \frac{1}{t_{l-1,l}}\left[(t_{l-2,l-1}-E)\psi_{l-1}-t_{l-2,l-1}\psi_{l-2}\right]\, ,\label{recursionDM}
\end{equation}
the main contribution to $\psi_l$ is for large 
$1/t_{l-1,l} \simeq \lambda $ 
approximately 
$$\psi_l \simeq \lambda t_{l-2,l-1}(\psi_{l-1}-\psi_{l-2})\,$$
which means that the wave function increases its amplitude by a factor of the order $\lambda$ during this step of the iteration, compared to how it would change in the absence of disorder. In contrast, a particularly small coupling strength $t_{l-2,l-1} \propto 1/\lambda$ (with a 'normal' value of $t_{l-1,l}$) does not reduce $\psi_l$ by a factor $\lambda$, since the recursion relation Eq.~\eqref{recursionDM} is then dominated by the first term, and $\psi_l \simeq \psi_{l-1}$. The wave function amplitude therefore gains an additional factor of the order of $\lambda$ during one wavelength, and so does 
 the area $S_N$. This means that the memory of the amplitude of the previous oscillation is lost over the distance of the order of one wavelength, implying that the localisation length is shortened from $\lambda^2$ to $\lambda$ when $w=2$. 
 
The power laws visible for $w=2$ in Figs.~\ref{fig:DOS} and \ref{fig:XiE_DM_different_w}(b) show
a small deviation from the exponent $-0.5$ expected based on this simple reasoning. We ascribe this to the fact that the wavelength of the modes is shortened by the disorder compared to what it would be without disorder; see Fig.~\ref{fig:DOS}(d). 

\subsection{Superdiffusive behaviour in the DM}
Our results for the DM also provide an intuitive explanation for the superdiffusive behaviour reported by Dunlap et al.~\cite{dunlap_absence_1989} for the case $w<2$.
Consider an eigenmode of wave length $\lambda$, which has an extension $\propto \lambda^2$. Such an eigenmode represents an unequal distribution of substance over a distance $\propto\lambda^2$. Its relaxation time is $\tau \propto \lambda^2$ (see Eq.~\eqref{DM2} where $\tau$ corresponds to $1/E$).
In the absence of disorder, relaxation of an eigenmode is achieved by a diffusive transport of the out-of-equilibrium substance. In the presence of disorder, part of this substance must be transported over the entire extension of the mode, i.e., over a distance $\propto\lambda^2$ during a time $\propto\lambda^2$. The proportion of substance that is transported over this distance during relaxation of the mode can be estimated from the difference in the amount of substance in the left and right half of the mode. Due to the random walk nature of the amplitude of the eigenmode this proportion is $\propto \frac{\sqrt{\lambda^2}}{\lambda^2} \propto 1/\lambda$. Taken together, the mean squared distance over which substance is transported during relaxation is 
$$\Delta x^2 \propto \frac 1 \lambda  \cdot \lambda^4 \propto \lambda^3 \propto \tau^{3/2}\, ,$$ which dominates the subleading diffusive term $\propto \tau$ and agrees with the exponent obtained by Dunlap et al.~\cite{dunlap_absence_1989}. 

\subsection{Logarithmic divergence of the localisation length as a function of $E$ in the RCM}
The transfer matrix for 2 iterations is given by 
\begin{align}
    \begin{pmatrix}
        \psi_{n+2}\\
        \psi_{n+1}
    \end{pmatrix} &= 
    \begin{pmatrix}
        \frac{E^2-t_{n,n+1}^2}{t_{n,n+1}t_{n+1,n+2}} & \frac{Et_{n-1,n}}{t_{n,n+1}t_{n+1,n+2}}\\
        -\frac{E}{t_{n,n+1}} & -\frac{t_{n-1,n}}{t_{n,n+1}}
    \end{pmatrix}
    \cdot\begin{pmatrix}
        \psi_n\\
        \psi_{n-1}
    \end{pmatrix} \label{eqTM2}
  \end{align}
  For small $E$, the wave functions of the two sublattices are only weakly coupled by a term of the order of $E$. If we ignore this coupling, $\ln{\psi}$ performs a random walk on each sublattice. As long as $\ln \psi$ is of the same order on the two sublattices the corrections due to the coupling are small. However, when the ratio of the wave functions on the two sublattices becomes of the order $1/E$, the evolution of the wave function with the smaller amplitude becomes dependent on that of the other sublattice. Since the distance between the logarithms of the wave functions performs a random walk, this happens on average after of the order of $(\ln(1/E))^2$ iterations. While the smaller wave function is coupled to the larger one,  $|\ln \psi|$ increases on average by an amount of the order of $\ln(1/E)$ until the coupling to the larger wave function becomes negligible again. On average, it will take another $(\ln(1/E))^2$ iterations until the two random walks meet again. 
  From these considerations follows that during $L$ iterations the two random walks become coupled $\sim L/(\ln (1/E))^2$ times, and $(\ln \psi)$ increases on average by an amount $\sim \ln(1/E) \cdot L/(\ln (1/E))^2 = L/\ln(1/E)$, resulting in a localisation length $\xi \sim \ln(1/E)$. Fig.~\ref{fig:TMM_examples}(b) illustrates this description of the behaviour of $\ln \psi$. The typical height of bubbles between the $\psi$ of the two sublattices is of the order of $-\ln E \simeq 16$, and the typical length of bubbles is of the order $(\ln E)^2 \simeq 260$.

The eigenmodes don't show the logarithmic increase of the localisation length (now measured as participation ratio) with $1/E$, as the participation ratio is small and independent of $E$ for small $E$, see Fig.~\ref{fig:EvP_HC}(a) and (c). We ascribe this to the exponential size of the wave function maxima, which include only a small number of lattice sites that contribute considerably to the participation ratio. The participation ratio therefore remains finite as $E$ goes to 0. (This is different when $\ln \psi$ instead of $\psi$ is used when calculating the participation ratio and the localisation length. The random walk nature of the logarithm of the wave function gives a localisation length and a participation ratio that increase as $L$ at $E=0$, apart from logarithmic corrections. We confirmed this by constructing for different $L$ ensembles of systems that have eigenstates at $E=0$, and by evaluating the participation ratio of $\ln \psi$ for $E=0$, but did not include those plots.)

\section{Summary and Conclusions}

We have investigated two one-dimensional tight-binding models with uncorrelated disorder and nearest-neighbor hopping that have a localisation length (as calculated with the transfer-matrix method) that diverges as the energy approaches zero. In the random-coupling model (RCM), disorder resides only in the off-diagonal terms of the Hamiltonian, while in the diffusion model (DM) the sum of the entries in each line of the Hamiltonian vanishes. In both models, the off-diagonal elements were chosen randomly from the interval $[1-w/2,1+w/2]$, with the disorder strength $w\in [0,2]$ being a parameter of the model. These two models are in contrast to the Anderson model, where the localisation length remains finite. In the RCM, the diverging localisation length is due to a system-spanning mode at $E=0$, where the two sublattices decouple and the logarithm of the wave function of each sublattice performs an independent random walk. For small $E$, the two random walks become weakly coupled, and by exploiting this feature we could give an intuitive explanation of the relation $\xi \sim \ln(1/E)$. Interestingly, the participation ratio of the eigenmodes of the RCM does not show this logarithmic increase, but remains finite in the limit $E \to 0$. Since the participation ratio $P$ is calculated directly from the wave function amplitudes (and not from the logarithm, as the localisation length), the exponentially sharp peaks dominate the evaluation of $P$.

In the DM, the localisation length and the participation ratio $P$ both diverge as $1/E$. The dispersion relation, the wavelength of the wave functions, and the relation between $P$ and $E$ show also power laws, for all of which we could provide intuitive explanations. From these we derived that the number of system-spanning modes scales with the system size $L$ as $\sqrt L$. These system-spanning modes are due to the conservation law of the system, and they are sine functions like those of the ordered system ($w=0$). Furthermore, we explored the behaviour of the model as the disorder strength $w$ approaches the value 2, so that the smallest hopping terms can be arbitrarily close to zero. We showed that the power laws become different for $w=2$ and that there is a crossover between the two scaling regimes, with the crossover energy being proportional to $(2-w)^2$ and the crossover length scale being proportional to $1/(2-w)$. Again, we could provide an intuitive explanation. The RCM does not show a qualitative change as $w$ approaches 2, but the slope $a$ of the law $\xi = -a\ln E$ changes continuously with $w$. In higher dimensions, we do not expect any more a qualitative  difference between $w=2$ and $w<2$ in the DM, since hopping terms close to zero do not block any more the flow of substance through the system.

The diverging localisation lengths in the two models do not contradict Furstenberg's theorem \cite{furstenberg1960products} that the norm of the product of transfer matrices shows an exponential increase (implying a finite localisation length) for almost every realization. In our models, the fraction of system-spanning modes goes to zero as the system size goes to infinity, so that almost all modes become localised. Similarly, the scaling theory of localisation \cite{abrahams_scaling_1979} makes a generic statement about localisation in one dimension that does not rule out the existence of a vanishing fraction of system-spanning modes. 

The system-spanning modes at $E=0$ will occur in both models also in higher dimensions for strong disorder, when almost all modes are localized. We have argued that in the DM the conserved quantity is responsible for the system-spanning modes. For the RCM, a bipartite structure of the lattice leads to a systems-spanning mode in the band center \cite{eilmesExponentsLocalizationLengths2001a,inuiUnusualPropertiesMidband1994}. 

Despite all the insights provided by our investigation, there remain open questions: Our data for the DM show that for $w=2$ the wavelength of the modes does not scale exactly  as $1/\sqrt E$ anymore, but is shorter. We did not perform analytical calculations to evaluate this deviation quantitatively. Neither did we succeed at obtaining a quantitative estimate of the typical difference of the highest peaks of the wave functions on the two sublattices of the RCM, which, however, is important for calculating the asymptotic value of the participation ratio as $E \to 0$. Also, the case $w=2$ for the RCM would deserve more attention, as Fig.~2(c) shows that the smallest and largest values of the participation ratio scatter considerably. 

Our paper thus shows that even models that have been studied for decades still conceal surprising features and deserve being revisited. The vastly increased numerical power of computers aids these studies and allowed us to obtain a more comprehensive understanding of the models based on the eigenmodes of the systems. 

\subsection*{Acknowledgements}
We acknowledge financial support for this project by DFG grant number Dr300/16.

\printbibliography
\end{document}